# Space-Varying Iterative Restoration of 2-D Inversion Models Computed from Marine CSEM Data

Feng-Ping Li [1,2], Vemund Stenbekk Thorkildsen [2], Leiv-J Gelius [2], Jian-Hua Yue[1]

[1]Department of Geophysics, China University of Mining and Technology, Xuzhou, China

[2]Department of Geosciences, University of Oslo, Norway

## ABSTRACT

Marine Controlled Source Electromagnetic (CSEM) is employed both in large-scale geophysical applications as well as within exploration of hydrocarbons and gas hydrates. Due to the diffusive character of the EM field only very low frequencies are used leading to inversion results with rather low resolution. In this paper, we calculate the resolution matrix associated with the inversion and derive the corresponding point spread functions (PSFs). The PSFs give information about how much the actual inversion has been blurred, and use of space-varying deconvolution can therefore further improve the inversion result. The actual deblurring is carried out by use of the nonnegative flexible conjugate gradient algorithm for least squares problem (NN-FCGLS), which is a fast iterative restoration technique. For completeness, we also introduce results obtained by use of a blind deconvolution algorithm based on maximum likelihood estimation (MLE) with unknown PSFs. The potential of the proposed approaches have been demonstrated using both complex synthetic data as well as field data acquired at the Wisting oil field in the Barents Sea. In both cases, the resolution of the final inversion result has improved and shows better agreement with the known target area.

## 1 Introduction

The Marine Controlled Source Electromagnetic technique has been proved successful in the search for thin hydrocarbon and natural gas reservoirs with high resistivity in seabed deposits. Thus, when integrated with seismic data, an improved image of the fluid distribution in a reservoir can be achieved (Um & Alumbaugh, 2007; Constable, 2010).

In parallel with the development of CSEM from an equipment point of view, major advances were also achieved regarding processing and interpretation of the acquired data. Initially this analysis was carried out in the data domain by the use of normalized magnitude versus offset (MVO) plots (Ellingsrud et al., 2002; Røsten et al., 2003). However, as the computing power developed, a complete inversion/imaging in the model domain replaced the simple data domain approach. These days, the reconstruction techniques allow the inference of finely discretized and anisotropic earth model estimates usually presented as 2-D or 3-D images (Brown et al., 2012; Wang et al., 2018; Jakobsen & Tveit, 2018).

The geophysical inverse problems always involve the search for an earth model consistent with only a limited amount of observations. However, data noise, bias, and use of inappropriate priori geological information, all lead to uncertainties in the solution model. Thus, an infinite resolving power is never achieved. From a practical point of view, we therefore need a method that can estimate the resolving power of a given set of limited observations. An effective strategy is to make use of point-spread functions (PSFs) or Backus–Gilbert averaging kernels (Backus & Gilbert, 1968, 1970). These quantities can be extracted from the model resolution matrix (Jackson, 1972; Menke, 1984, 1989, 2012), which provides a useful tool for quantitative resolution analysis. Such an approach has been used to facilitate survey design based on both linearized approximations (Torres–Verdin, 1991; Zhou et al., 1993; Spies and Habashy, 1995) and nonlinear inversion model studies (Alumbaugh and Morrison, 1995). Additional studies have been carried out to perform a posteriori appraisal of seismic tomography, gravity, dc resistivity and electromagnetic inversion results, to determine the resolution of the inverted model and to estimate its error (Jackson, 1972, 1979; Parker, 1980; Oldenburg, 1983; Dosso and Oldenburg, 1989, 1991; Sen et al., 1993; Gouveia & Scales, 1997; Alumbaugh & Newman, 2000; Berryman, 2000; Friedel, 2003; Günther, 2004; Routh, 2005; Routh & Miller , 2006; Gao et al., 2007; Xia et al., 2008; Oldenborger & Routh, 2009; Routh, 2009; An, 2012; Mordret et al., 2013; Kalscheuer et al., 2013; Grayver et al., 2014; Fichtner & Leeuwen, 2015; Mattsson, 2015; Bogiatzis et al., 2016; Chrapkiewicz et al., 2020; Ren & Kalscheuer, 2020).

Thus, in the literature, the use of the resolution matrix has focused on quantitative interpretation, deriving the upper and lower limits of model parameters, optimizing survey design, and evaluation of inversion results. The general structure of a model resolution matrix is shown in figure 1 and further described by Menke (1984). Due to the diffusive character of the EM field only very low frequencies are used in marine CSEM leading to inversion results with rather low resolution. We propose in this work to extract the PSFs computed from a regularized inversion of marine CSEM data and then employ this information to deblur the inversion result by the use of deconvolution.

Several examples can be found in the literature using deconvolution to enhance the quality of geophysical techniques. Predictive deconvolution proposed by Peacock & Treitel (1969) allows the length of the desired output wavelet to be controlled, thus specifying the desired resolution. Reid (1990) demonstrated a fast way of processing magnetic-survey data in grid form for source positions and depths by Euler deconvolution. Li et al. (1995, 1997) proposed a pure phase-shift filter to eliminate the correlation-ghost sweeps by using deterministic deconvolution instead of the conventional correlation of vibroseis traces. Keating (1998) demonstrated the possibility of rejecting solutions resulting from aliasing and less accurate measurements by weighted Euler deconvolution for rapid interpretation of gravity data. Sjoeberg et al. (2003) introduced a 2-D deconvolution technique to improve the seismic image quality based on the knowledge of the blur-functions. Zuo & Hu (2012) employed a blind deconvolution to reduce errors that inevitably occur during Magneto Telluric (MT) data acquisition and subsequent processing. Wang et al. (2016) developed a novel sparse-spike deconvolution (SSD) method based on Toeplitz-sparse matrix factorization to effectively derive the wavelet and reflectivity simultaneously from band-limited data with appropriate lateral coherency. Kazemi (2018) provided an efficient and reliable single-channel blind-deconvolution technique to recover the reflectivity series and the wavelet without compromising the small amplitude events in the case of seismic recordings with a high signal-to-noise ratio. Yang et al. (2022) introduced an efficient and stable high-resolution seismic imaging method, in which the PSFs were used as a filter for deconvolution, thereby the effect of the band-limited source function is reduced and the irregular subsurface illumination is compensated. However, to our knowledge, the implementation of image restoration with the use of PSFs for inversion models computed from marine CSEM data has not yet been published.

Here, we propose to enhance the resolution, based on deconvolution as a post-processing step of the 2-D inversion models. The loss of resolution manifests itself as a blurring of the inversion results. Such blurring can be described by point-spread functions (PSFs) that vary spatially. Therefore, a space-varying blur model should be applied for restoration of marine CSEM inversions. The formulation developed by Nagy et al. (1997, 1998, 2004) is adopted here. It assumes that the PSF is spatially invariant in small regions of the image domain. In order to form a matrix for deconvolution, the invariant PSFs corresponding to sub-regions are stitched together by interpolation. Then a standard iterative algorithm for solving linear algebraic equations is used to restore the blurred image.

This paper is organized as follows. First, the marine CSEM inversion scheme and construction of the resolution matrices are briefly discussed. Next, the space-varying blur model is introduced followed by a discussion on how to restore the blurred image by use of the nonnegative flexible conjugate gradient algorithm for least-squares problems (NN-FCGLS). In addition, we have also considered blind deconvolution based on maximum likelihood estimation for deblurring in case of unknown PSFs.

Finally, the feasibility of the proposed approaches are demonstrated by use of both complex synthetic data as well as field data from the Wisting oil field in the Barents Sea.

## 2 THEORETICAL CONCEPTS OF INVERSION AND RESOLUTION MATRICES

### 2.1 Inversion framework

The 2-D regularized inversion scheme employed in this study is described by Key (2016), and has been implemented in the open-source inversion package MARE2DEM (Modelling with Adaptively Refining Elements 2D EM). MARE2DEM is developed for 2D anisotropic modeling and inversion of controlled-source electromagnetic (CSEM), Magneto Telluric (MT) and surface-borehole EM applications in onshore, offshore and downhole environments. This package adopts a fast Occam approach, and its original form is introduced by Constable et al. (1987). A nonlinear inverse problem is generally solved iteratively by minimizing a cost function as follows (Ren & Kalscheuer, 2020):

$$U[\mathbf{m},\alpha] = Q_d[\mathbf{m}] + \alpha Q_m[\mathbf{m}] \quad , \tag{1}$$

where $\mathbf{m}$ is the model vector, $Q_d[\mathbf{m}]$ is the data misfit depending on the model, and $Q_m[\mathbf{m}]$ is the regularization term employed to simplify the solution space. The Lagrangian multiplier $\alpha$ is used as a weighting factor between the data misfit term and the regularization term, so as to balance resolution and stability. In most cases, we assume that the uncertainty $\delta_i$, $i = 1, \ldots, N$ in each observation data originates from random noise with a normally distributed probability of zero mean and non-zero standard deviation. Equation 1 can be expanded as

$$U[\mathbf{m},\alpha] = \left[ (\mathbf{d} - \mathbf{F}[\mathbf{m}])^\dagger \mathbf{W_d}^\dagger \mathbf{W_d} (\mathbf{d} - \mathbf{F}[\mathbf{m}]) \right] + \alpha \mathbf{m}^\dagger \mathbf{W_m}^\dagger \mathbf{W_m} \mathbf{m}, \tag{2}$$

where $\mathbf{d}$ of size N represents the measured complex field data and $\mathbf{F}[\mathbf{m}]$ the corresponding model response. While dealing with complex fields, we need to adopt the Hermitian † (i.e matrix transpose + complex conjugation) notation for the matrices involved. The data misfit is also weighted by the matrix $\mathbf{W_d}$, $W_d^{ii} = 1/\delta_i$, which is a diagonal matrix composed of the reciprocal of the standard error for each sample. The regularization term includes the weighting matrix $\mathbf{W_m}$ that forces the smoothness of the model. With MARE2DEM, this is obtained by using a gradient roughness operator. As for the anisotropic models, the roughness is enlarged by partitioning the model vector into anisotropic subsets (Key, 2016).

Due to the non-linearity of the inverse problem, the forward operator is quasi-linearized by use of a Taylor series expansion. Thus, the parameter update at the $(k+1)^{th}$ iteration is given by:

$$U^{lin}[\mathbf{m}_{k+1}, \alpha] = \left[ \left( \mathbf{d} - \mathbf{F}[\mathbf{m}_k] - \mathbf{J}(\mathbf{m}_{k+1} - \mathbf{m}_k) \right)^\dagger \mathbf{W}_\mathbf{d}^\dagger \mathbf{W}_\mathbf{d} \left( \mathbf{d} - \mathbf{F}[\mathbf{m}_k] - \mathbf{J}(\mathbf{m}_{k+1} - \mathbf{m}_k) \right) \right] \\ + \alpha \mathbf{m}_{k+1}^\dagger \mathbf{W}_\mathbf{m}^\dagger \mathbf{W}_\mathbf{m} \mathbf{m}_{k+1}. \qquad (3)$$

The Jacobian or sensitivity matrix J (with entries $\partial F_i(m_k)/\partial \log(\rho_j)$, where $\rho_j$ is the resistivity in cell $j$) consists of the first order partial derivatives with respect to model parameters (log resistivity). By setting $\partial U^{lin}[\mathbf{m}_{k+1}, \alpha]/\partial m_{k+1}$ to zero and solving for $\mathbf{m}_{k+1}$, a least squares solution is derived:

$$\mathbf{m}_{k+1} = \mathbf{J}_w^{-g} \mathbf{W}_\mathbf{d} \mathbf{d}_k, \qquad (4)$$

where $\mathbf{d}_k = [\mathbf{d} - \mathbf{F}[\mathbf{m}_k] + \mathbf{J}\mathbf{m}_k]$ and $\mathbf{J}_w^{-g}$ is the generalized inverse matrix $[\mathbf{J}^\dagger \mathbf{W}_d^\dagger \mathbf{W}_d \mathbf{J} + \alpha \mathbf{W}_m^\dagger \mathbf{W}_m]^{-1} \mathbf{J}^\dagger \mathbf{W}_d^\dagger$. In general, for an EM problem, a total of six different data components exist, which are respectively related to the three different directions of the magnetic and electric field. In this study, however, we only use the embedded horizontal electric field ($E_y$), which is known to be the most important one in marine CSEM. For more details on the inversion procedure and how the components are calculated, the reader is referred to Key (2016).

## 2.2 Model resolution matrix

The *model resolution matrix*, $\mathbf{R}_\mathbf{M}$, defines the connection between the true model, $\mathbf{m}_{true}$, and the model ($\mathbf{m}_{k+1}$) estimated from non-linear least-squares inversion as follows (Jackson, 1972; Menke, 1984, 1989, 2012)

$$\mathbf{m}_{k+1} = \mathbf{R}_\mathbf{M} \mathbf{m}_{true} + \mathbf{J}_w^{-g} \mathbf{W}_d \mathbf{n}, \qquad (5)$$

where **n** represents noise, and $\mathbf{R}_\mathbf{M}$ is explicitly given as (Ren & Kalscheuer, 2020):

$$\mathbf{R}_\mathbf{M} = \Re\left[ \left[ \mathbf{J}^\dagger \mathbf{W}_d^\dagger \mathbf{W}_d \mathbf{J} + \alpha \mathbf{W}_m^\dagger \mathbf{W}_m \right]^{-1} \mathbf{J}^\dagger \mathbf{W}_d^\dagger \mathbf{W}_d \mathbf{J} \right], \qquad (6)$$

where $\Re$ indicates taking the real part. $\mathbf{m}_{k+1}$ is considered as the preferred inversion model if the inversion is terminated at the $k^{th}$ iteration. Note that equation 5 follows directly from equation 4 by setting (also allowing for measurement noise $\mathbf{n}$) $\mathbf{d}_k = \mathbf{d}_{true} = \mathbf{Jm}_{true} + \mathbf{n}$. Thus, in a practical inversion case $\mathbf{m}_{true}$ is unobtainable. The model resolution matrix reveals how close the preferred inversion model is to the true model, which relies on the Lagrangian multiplier $\alpha$. By letting $\alpha \to 0$, the model resolution matrix will approximate the unity matrix. In this case, it is said that the inverse problem is perfectly solved if no noise.

By decomposing $\mathbf{R}_M$ into its column, the model resolution matrix $\mathbf{R}_M$ can be written as

$$\mathbf{R}_M = [\mathbf{r}_1 \cdots \mathbf{r}_k \cdots \mathbf{r}_M], \qquad (7)$$

where $\mathbf{r}_i$ denotes the *i*-th column vector, and with $i = 1, \ldots, M$. The *model resolution matrix* contains information about how the unavailable true model is actually retrieved by the inversion scheme (i.e. the blurring). Figure 1a illustrates the information content of equation 7, where each column represents a Point Spread Function (PSF). As is well known from imaging theory, the PSF describes how the imaging system responds to impulses. If a delta function is specified in $\mathbf{m}_{true}$, the PSF describes how this delta function spreads across the inverted model $\mathbf{m}_{k+1}$ (Figure 1a). Each model parameter has its own unique PSF, represented by the columns of $\mathbf{R}_M$. Therefore, if we want to analyze the PSF of any point in the image domain, we can do it simply by plotting the appropriate column of the $\mathbf{R}_M$ after reorganizing from a 1D to a 2D model parameter representation (Figures 1b and 1c). Ideally, if the model is perfectly determined, the associated PSFs are delta functions, and the resolution matrix is equal to the unity matrix. Most of the time, such a model is unlikely to be obtained, and the PSFs will vary across the model space. In the usual inversion, however, it is likely that some areas will be well recovered, while others will be poorly distinguished. In a well-resolved region, the PSFs will be characterized by small spread centered on the corresponding model parameters (Figure 1b). A point spread function in a region with poor resolution can be characterized by a large spread, an off-center maximum or a combination of the two (Figure1c).

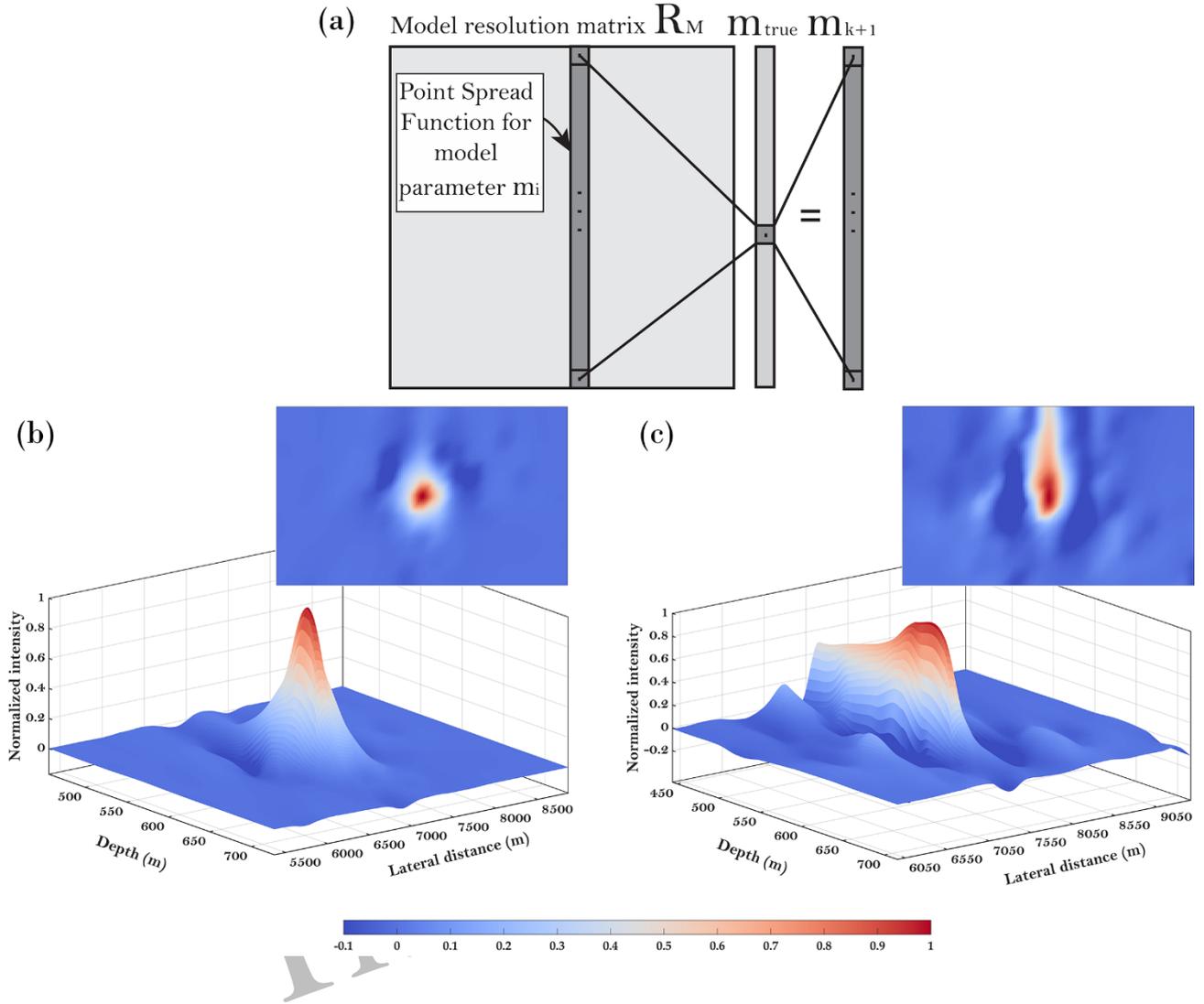

**Figure 1:** *(a) The relationship between the unobtainable true model $\mathbf{m}_{true}$ and the preferred inversion model $\mathbf{m}_{k+1}$ expressed by the PSF. (b) and (c) are two interpolated point spread functions determined from the model resolution matrix for the image given in Figure 10. For presentation purposes the results have been normalized such that the maximum value in each case is equal to unity.*

## 3 DEBLURRING OF INVERSION RESULT

### 3.1 Space-varying blur model

To set the stage and to introduce basic notation, we start with a brief review of the general relationship between the actual blurred image (from inversion) and the corresponding true image. The corresponding mathematical model can be formulated as

$$\mathbf{b}=\mathbf{A}\mathbf{m}_{true}+\boldsymbol{\varepsilon}, \tag{8}$$

where **b** is the blurred, noisy image (i.e. actual inversion), $\mathbf{m}_{true}$ is the unknown true image, **ε** is additive noise, and **A** is the blurring matrix.

In 2D EM inversion, the PSFs may vary spatially along both directions (Alumbaugh, 2000). The representation of a generic spatially variant blur would require a PSF attached at every pixel location to fully describe the blurring operation. This is not computationally feasible, even for small images, and some approximations should be introduced. Several approaches to restore images degraded by a spatially variant blur exist. In this study, we adapted the blur model developed by Nagy et al. (1997, 1998, 2004). The blurred image is then partitioned into sub-regions of Figure 2 with each of them assigned a spatially invariant PSF. However, rather than deblurring each individual sub-region and then sewing the individual results together, this method interpolates the individual PSFs, and restores the image globally.

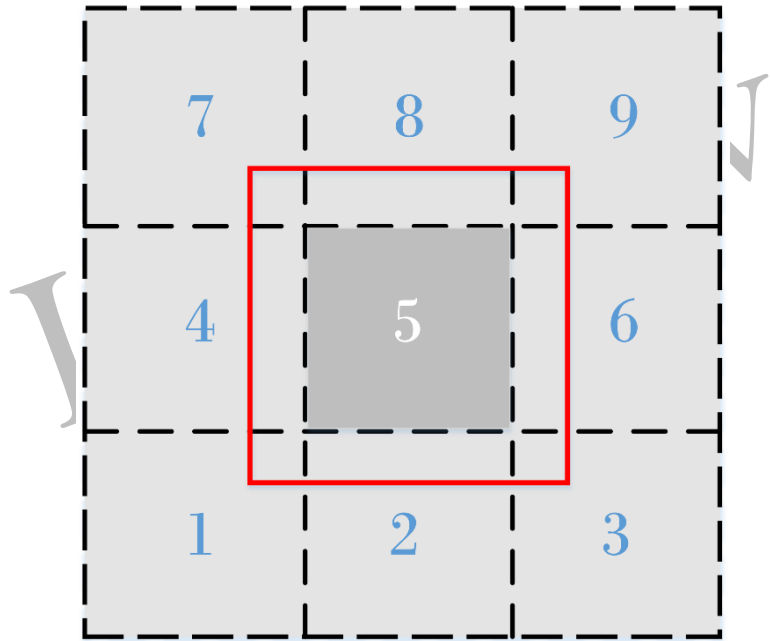

**Figure 2:** *Example of image domain segmentation (area within the black dashed box). The domain section within the red solid frame for sub-region 5 is overlaid to show the overlap of the PSF domain.*

Assume the N×N image is partitioned into $p \times q$ sub-regions, and define the dimensions of the (*i, j*) th region as $m_i \times n_j$, where $m_1 + \cdots + m_p = n_1 + \cdots + n_q = N$. In addition to the boundary problem (discussed in the next subsection), another problem arises from the fact that different PSFs are associated with adjacent parts. Accordingly, even if we consider the case of slowly changing PSFs, deconvolution of disjoint domains will cause discontinuities at the shared boundaries. Camera et al.

(2015) introduced a procedure to extend each non-overlapping section corresponding to different PSFs, to an appropriate broader section sharing the same PSF. Therefore, we extend the disjoint part to the partially overlapping section. As a illustration, we have segmented the image domain into 3×3 rectangular blocks (Figure 2). Suppose, the domain section within the red solid frame for sub-region 5 has a pixel dimension of $(2\beta+1) \times (2\beta+1)$. Thus the blurring matrix $\mathbf{A}$ and the blurred matrix $\mathbf{b}_{ij}$ for the overlapping sub-region $(i, j)$ have the following structure:

$$\mathbf{A} = \sum_{i=1}^{p}\sum_{j=1}^{q} \mathbf{D}_{ij}\mathbf{A}_{ij}, \tag{9}$$

$$\mathbf{b}_{ij} = \begin{pmatrix} b_{-\beta-\beta} & \cdots & b_{-\beta-1} & b_{-\beta 0} & b_{-\beta 1} & \cdots & b_{-\beta\beta} \\ \vdots & \ddots & \ddots & \ddots & \ddots & \ddots & \vdots \\ b_{-1-\beta} & \ddots & b_{-1-1} & b_{-10} & b_{-11} & \ddots & b_{-1\beta} \\ b_{0-\beta} & \ddots & b_{0-1} & b_{00} & b_{01} & \ddots & b_{0\beta} \\ b_{1-\beta} & \ddots & b_{1-1} & b_{10} & b_{11} & \ddots & b_{1\beta} \\ \vdots & \ddots & \ddots & \ddots & \ddots & \ddots & \vdots \\ b_{\beta-\beta} & \cdots & b_{\beta-1} & b_{\beta 0} & b_{\beta 1} & \cdots & b_{\beta\beta} \end{pmatrix} \tag{10}$$

where $\mathbf{D}_{ij}$ is a non-negative diagonal matrix satisfying the condition $\sum_{i=1}^{p}\sum_{j=1}^{q} \mathbf{D}_{ij} = \mathbf{I}$, and $\mathbf{I}$ is the identity matrix. Moreover, $\mathbf{A}_{ij}$ is a block Toeplitz matrix with Toeplitz blocks (BTTB) defined by an invariant PSF for each overlapping sub-region $(i, j)$, and with zero boundary conditions (Nagy et al., 1998, 2003), this matrix can be written on the form

$$\mathbf{A}_{ij} = \begin{pmatrix} \mathbf{P}_0 & \mathbf{P}_{-1} & \cdots & \mathbf{P}_{-\beta} & 0 & \cdots & 0 \\ \mathbf{P}_1 & \ddots & \ddots & \ddots & \ddots & \ddots & \vdots \\ \vdots & \ddots & \mathbf{P}_0 & \mathbf{P}_{-1} & \ddots & \mathbf{P}_{-\beta} & 0 \\ \mathbf{P}_\beta & \ddots & \mathbf{P}_1 & \ddots & \mathbf{P}_{-1} & \ddots & \mathbf{P}_{-\beta} \\ 0 & \ddots & \ddots & \mathbf{P}_1 & \mathbf{P}_0 & \ddots & \vdots \\ \vdots & \ddots & \mathbf{P}_\beta & \ddots & \ddots & \ddots & \mathbf{P}_{-1} \\ 0 & \cdots & 0 & \mathbf{P}_\beta & \cdots & \mathbf{P}_1 & \mathbf{P}_0 \end{pmatrix} \tag{11}$$

where vector $\mathbf{P}_k$ is a $(2\beta+1) \times (2\beta+1)$ Toeplitz matrix (with $k = -\beta,...,-1,0,1,...,\beta$), which is constructed from the column vector representation of the invariant PSF ($\mathbf{PSF}_{ij}$) corresponding to the overlapping sub-region $(i,j)$ as follows:

$$\mathbf{P}_k = \text{Toeplitz}(\mathbf{PSF}_{ij}(:,k)) = \begin{pmatrix} p_{0k} & p_{-1k} & \cdots & p_{-\beta k} & 0 & \cdots & 0 \\ p_{1k} & \ddots & \ddots & \ddots & \ddots & \ddots & \vdots \\ \vdots & \ddots & p_{0k} & p_{-1k} & \ddots & p_{-\beta k} & 0 \\ p_{\beta k} & \ddots & p_{1k} & \ddots & p_{-1k} & \ddots & p_{-\beta k} \\ 0 & \ddots & \ddots & p_{1k} & p_{0k} & \ddots & \vdots \\ \vdots & \ddots & \ddots & \ddots & \ddots & \ddots & p_{-1k} \\ 0 & \cdots & 0 & p_{\beta k} & \cdots & p_{1k} & p_{0k} \end{pmatrix} \quad (12)$$

where the operator "Toeplitz $(\cdot)$" transforms the column vector of the $\mathbf{PSF}_{ij}$ matrix into a Toeplitz matrix. Therefore, the matrix-vector multiplication in Equation (9) involving the PSF can be performed by use of the 2D discrete fast Fourier transform (Nagy et al., 1997, 1998).

### 3.2 Sectioning of the image domain

The simplest form for $\mathbf{D}_{ij}$ is calculated by piecewise constant interpolation, where the allowed values are either 0 or 1. On this occasion, the non-zero entries compose the segmentation of the image domain. Although this option provides desirable results in many cases, sometimes the reconstruction results will be affected by block artifacts that cause problems for deblurring operations. The more complex alternative method (Nagy & O'Leary, 1998) that can get rid of such problems is based on linear interpolation; Vio et al. (2003b, 2005) use a modified Hanning window approach to deal with the edge effects, their results show that the edge effect in deblurring operations is significantly reduced; Camera et al. (2015) adopted a deconvolution method with boundary effects correction proposed in Bertero &

Boccacci (2005), and Anconelli et al. (2006) for each new overlapping profile. In this paper, we only segment the image equally according to the number of sampled PSFs, and then adopted the methodology of Nagy & O'Leary to restore the image globally.

### 3.3 Restoration algorithms

To restore the blurred image, we adopt the nonnegative flexible conjugate gradient algorithm for least squares problems (**NN-FCGLS**) associated with the semi–convergence case (Gazzola et al. , 2017), it solves the following minimization problem

$$\Phi_\alpha(\mathbf{m}) = \arg\min_{\alpha,\mathbf{m}} \|\mathbf{b} - \mathbf{Am}\|_2^2 \text{ s.t. } \mathbf{m} \geq 0 \tag{13}$$

where $\mathbf{A} \in \mathrm{R}^{M \times N}$, $\alpha$ refers to the bounded step length, and the constraint $\mathbf{m} \geq 0$ on $\mathbf{m} \in \mathrm{R}^N$ is intended component-wise. The parameter update at the $k$th iteration is given by

$$\mathbf{m}_k = \mathbf{m}_{k-1} + \alpha_{k-1} \mathbf{g}_{k-1} \tag{14}$$

where $\mathbf{g}_{k-1}$ is a new search direction added at each iteration. An iterative method applied as a regularization technique may give solutions with negative terms, however, this NN-FCGLS algorithm enforces a nonnegativity constraint on the solution approximation at each iteration. Such enforcing produce a much more accurate approximate solution in many practical cases of nonnegative true images, and this is the main reason why we choose it for our image restoration problems. The pseudo-code of NN-FCGLS method is given in the Table 1, note that the notations $\mathbf{m}_{k-1}(\mathbf{g}_{k-1} < 0)$ and $\mathbf{g}_{k-1}(\mathbf{g}_{k-1} < 0)$ indicate that only the elements of the vectors $\mathbf{m}_{k-1}$ and $\mathbf{g}_{k-1}$ corresponding to negative values of $\mathbf{g}_{k-1}$ are calculated, respectively. The reader is referred to Gazzola et al. (2017) for more details.

**Table 1. NonNegative FCGLS (NN-FCGLS) method.**

**Input**: $\mathbf{A}, \mathbf{b}, \mathbf{m}_0 \geq 0$ (initial model), $n_{\max}^{\text{in}}$, $n_{\max}^{\text{out}}$ (maxium iterations of respectively inner and outer loop).

$$\mathbf{L}_0 = \mathbf{M}_0 = \text{diag}(\mathbf{m}_0),$$

**For** $n = 1, \ldots,$ till a stopping criterion is satisfied **OR** $n = n_{\max}^{\text{out}}$ :

- $\mathbf{r}_{n-1} = \mathbf{b} - \mathbf{A}m_{n-1}, \mathbf{z}_{n-1} = \mathbf{A}^{\mathrm{T}}\mathbf{r}_{n-1}, \hat{\mathbf{z}}_{n-1} = \mathbf{L}_{n-1}\mathbf{z}_{n-1}, \mathbf{g}_{n-1} = \hat{\mathbf{z}}_{n-1}, \text{ and } w_{n-1} = \mathbf{A}\hat{\mathbf{z}}_{n-1}.$

- **For** $k = 1, \ldots,$ till a stopping criterion is satisfied **OR** $k = n_{\max}^{\text{in}}$ :

  Set $\alpha_{k-1} = (\mathbf{r}_{k-1}, w_{k-1}) / (w_{k-1}, w_{k-1})$ .

  $\hat{\alpha}_{k-1} = \mathbf{min}\left(\alpha_{k-1}, \min\left(-\mathbf{m}_{k-1}(\mathbf{g}_{k-1} < 0) / \mathbf{g}_{k-1}(\mathbf{g}_{k-1} < 0)\right)\right),$

  $\mathbf{m}_k = \mathbf{m}_{k-1} + \hat{\alpha}_{k-1}\mathbf{g}_{k-1}.$

  $\mathbf{L}_k = \mathbf{M}_k = \text{diag}(\mathbf{m}_k).$

  $r_k = r_{k-1} - \hat{\alpha}_{k-1} w_{k-1}.$

  $z_k = \mathbf{A}^{\mathrm{T}} r_k$ and $\hat{z}_k = L_k z_k$ .

  **For** $j = \max\{0, k - n_{\max}^{\text{in}}\}, \ldots, k-1,$

  $\beta_j = -(\mathbf{A}\hat{z}_k, w_j) / (w_j, w_j).$

  $g_k = \hat{z}_k + \sum_{j=\max\{0, k-n_{\max}^{\text{in}}\}}^{k-1} \beta_j g_j$ .

  $w_k = \mathbf{A}g_k = \mathbf{A}\hat{z}_k + \sum_{j=\max\{0, k-n_{\max}^{\text{in}}\}}^{k-1} \beta_j w_j$ .

  **end**

- **Until** $n_k$ (the stopping criterion) is reached, **then** $\mathbf{M} = \mathbf{m}_{n_k}$.

**end**

### 3.4 Stop criteria

The **NN-FCGLS** algorithm makes use of the discrepancy principle to stop in the most optimal iteration, which means that once the relative norm of the residual $\mathbf{b} - \mathbf{A}\mathbf{m}_k$ is sufficiently small, usually the same as the norm of the noise $\boldsymbol{\varepsilon}$, the algorithm stops, that is, when

$$\frac{\|\mathbf{b} - \mathbf{A}\mathbf{m}_k\|_2}{\|\mathbf{b}\|_2} \leq \eta \cdot \chi, \tag{15}$$

where $\eta$ is a "safety factor" that is slightly larger than 1, and $\chi = \|\boldsymbol{\varepsilon}\|_2 / \|\mathbf{b}\|_2$ is the relative noise level. The default value of 0 is used in the code if the noise level $\chi$ is not assigned. We may set $\eta = 1$ and $\chi = \tau$ to solve the noise-free problem of the given threshold $\tau$. Since the **NN-FCGLS** algorithm employs inner and outer iterations, the discrepancy principle is implemented for the solver in the inner iterations, and the outer iterations are terminated when either a stopping rule has been reached, the method stagnates or reached maximum number of (total) iterations.

## 3.5 Blind deconvolution

The above deconvolution method uses the calculated PSFs to restore the blurred image. Thus, the accuracy of this process is limited by the validity of the model resolution matrix used as a relative metric to evaluate the inversion results. Please note, however, that the nonlinear inverse problem in this paper is linearly approximated in the actual inversion process, and the estimated PSFs are not exact, so the obtained results should be treated with caution. The method discussed in this section does not use calculated PSFs, and is called blind deconvolution (Ayers and Dainty, 1988; Holmes, 1992; Krishnamurthi et al., 1995; Bhattacharyya et al., 1996). Therefore, when the equivalent blurring matrix $\mathbf{A}$ is unknown, it is necessary to use a deblurring technique that can jointly estimate PSFs and the unknown true image $\mathbf{m}_{true}$.

To solve the blind deconvolution problem, the Maximum Likelihood Estimation (**MLE**) principle (Van Trees, 1968) is employed. As the logic of maximum likelihood estimation is intuitive and flexible, this method has become the main means of statistical inference. Given some observed data (i.e. blurred matrix $\mathbf{b}$), it is realized by maximizing the likelihood function. Therefore, under the assumed statistical model, the parameter values (i.e. PSFs and $\mathbf{m}$) deduced are most likely (maximum probability) to lead to such observed data. For more details, the reader is referred to Holmes et al. (1992, 1995, 2006), Krishnamurthi et al. (1992), and Biggs and M. Andrews (1997). In this paper, we employ blind deconvolution to benchmark our proposed approach.

# 4 SYNTHETIC DATA EXAMPLE

## 4.1 Synthetic data inversion

To demonstrate the applicability of the deblurring techniques, we use the model shown in Figure 3a, which consists of faulted blocks and an oil-water contact (1500 Ohm-m, red components) embedded in a subsurface. This is a high-quality synthetic model of the subsurface, which was constructed based on the Wisting oil field in the South Western Barents Sea. Wisting is an oil field where CSEM has been verified to provide significant information. The geometric structure shown in Figure 3a is established by accessing resistivity logging data, high-quality CSEM and seismic field data from the same locality.

In the case of CSEM, assuming isotropic earth will lead to poor inversion results. Previous studies have shown that a moderate ratio of vertical to horizontal resistivity of 2-3 may have an appreciable effect on the inverted results. (Lu & Xia, 2007; Newman et al., 2010; Brown et al., 2012). Additionally, in the case of anisotropic earth, by adopting isotropic inversion scheme, the inversion will be biased towards higher resistivity value. This is because CSEM data is usually more sensitive to vertical resistivity (Hoversten et al., 2006).

Generally, the resistivity measured in the borehole is perpendicular to the shaft, which means that the resistivity obtained in the vertical borehole is a horizontal component. Therefore, in the case of acquisition in deviated wells, the obtained resistivity will be a composition of vertical and horizontal resistivity. The measured value of vertical resistivity in this project is beyond our reach. The operator of Wisting oil field (Equinor), however, provided the representative value of anisotropy. The final vertical resistivity values of all formations and the ratio of vertical resistivity to horizontal resistivity ( $\rho_z/\rho_{xy}$ ) are listed in Table 2, and a brief description of lithology is presented. In addition, all the typical anisotropic factors are found in the related literatures.

In the oil-bearing part of the reservoir (three red part in Figure 3a), there are three main strata, Stø stratum is the main hydrocarbon-bearing unit, while Nordmela and Fruholmen strata constitutes the remaining part of the reservoir (Granli et al., 2017). Synthetic marine CSEM data were generated using the model shown in Figure 3a with the MARE2DEM package, and its transmitters and receivers has the same number and relative position as the acquisition system of field data, and 27 frequencies, from 0.2Hz to 14.4Hz, were used, as well as random noise with a standard deviation equal to 1% of the data amplitude was added to each datum. The estimated model obtained by nonlinear inversion scheme is shown in Figure 3b.

**Table 2:** *Vertical and horizontal resistivity values of the Wisting model. Please note that Stø and Nordmela are listed twice, because oil reservoirs are formed in these two strata. The resistivity of oil-filled Stø formation is not given a definite value, but only a range is listed here. Moreover, the lithology description in the table is quoted from Senger et al. (2021).*

| Formation | $\rho_z$ [Ohm-m] | $\frac{\rho_z}{\rho_{xy}}$ | Lithology |
|---|---|---|---|
| Nordland GP | 7 | 2.3 | Marine Shale |
| Kollmule FM | 15 | 3.4 | Marine Shale |
| Kolje FM | 15 | 2.7 | Marine Shale |
| Hekkingen FM | 19.5 | 3.2 | Marine/organic rich shale |
| Fuglen FM | 19.5 | 2.4 | Marine Shale |
| Stø FM(oil filled) | 1500-2500 | 1 | Sandstone |
| Stø FM(brine filled) | 3 | 2 | Sandstone |
| Normela FM(oil filled) | 50 | 1 | Marine Shale/ Sandstone |
| Normela FM(brine filled) | 7 | 2 | Marine Shale/ Sandstone |
| Fruholmen FM | 10 | 2 | Alluvial Shale/ Sandstone |
| Snadd Fm | 30 | 2 | Marine Shale |

From the perspective of expected results, the highest value (related to good resolution) of the inversion result of the synthetic data for the Wisting model should be positioned inside the reservoir, while the areas above and below the reservoir are identified by lower values (except some boundary effects). From the inversion results shown in Figure 3b, it can be seen that the final inverted resistivity model is characterized by three main compartments. Compared directly with the real (synthetic) model (Figure 3a), the inversion has captured the main features of the reservoir, especially the lateral extension. As CSEM method is generally lower in resolution than seismic imaging, however, the output image is characteristically smeared on a large vertical area.

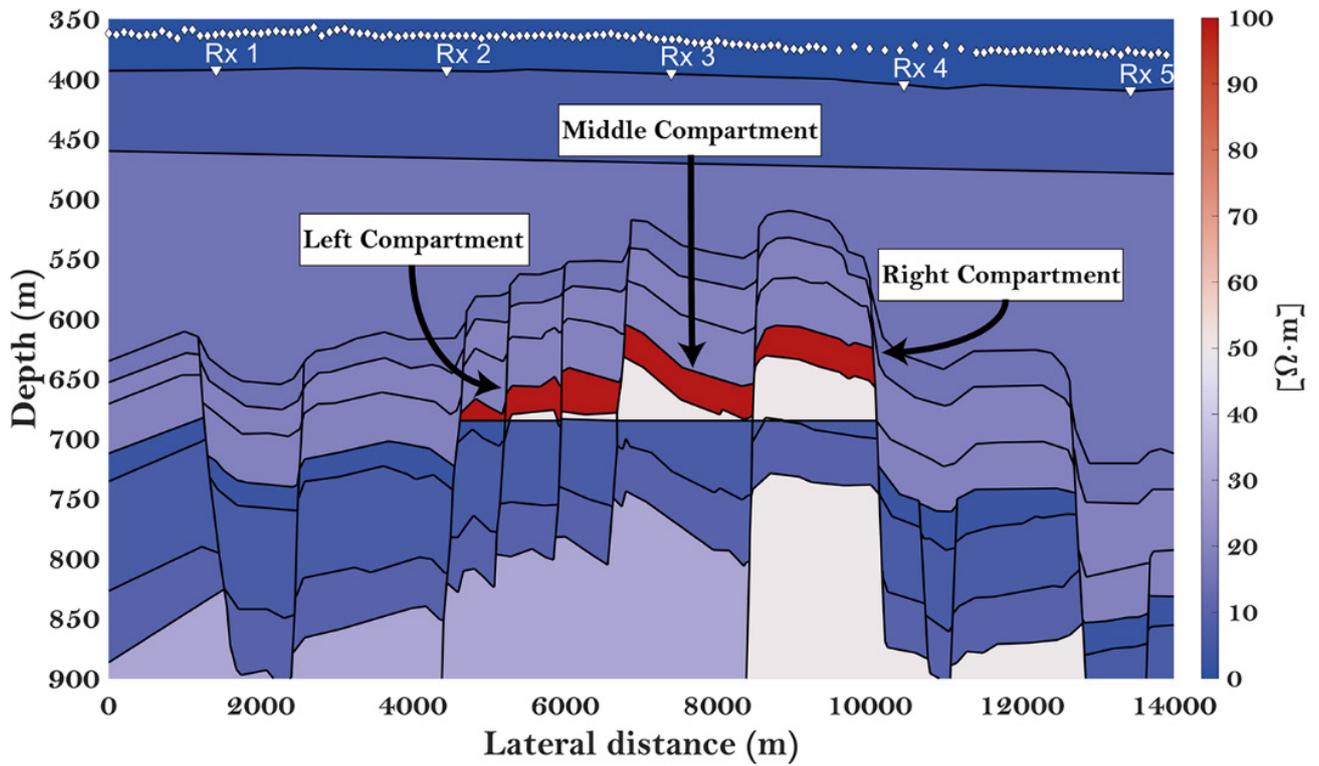

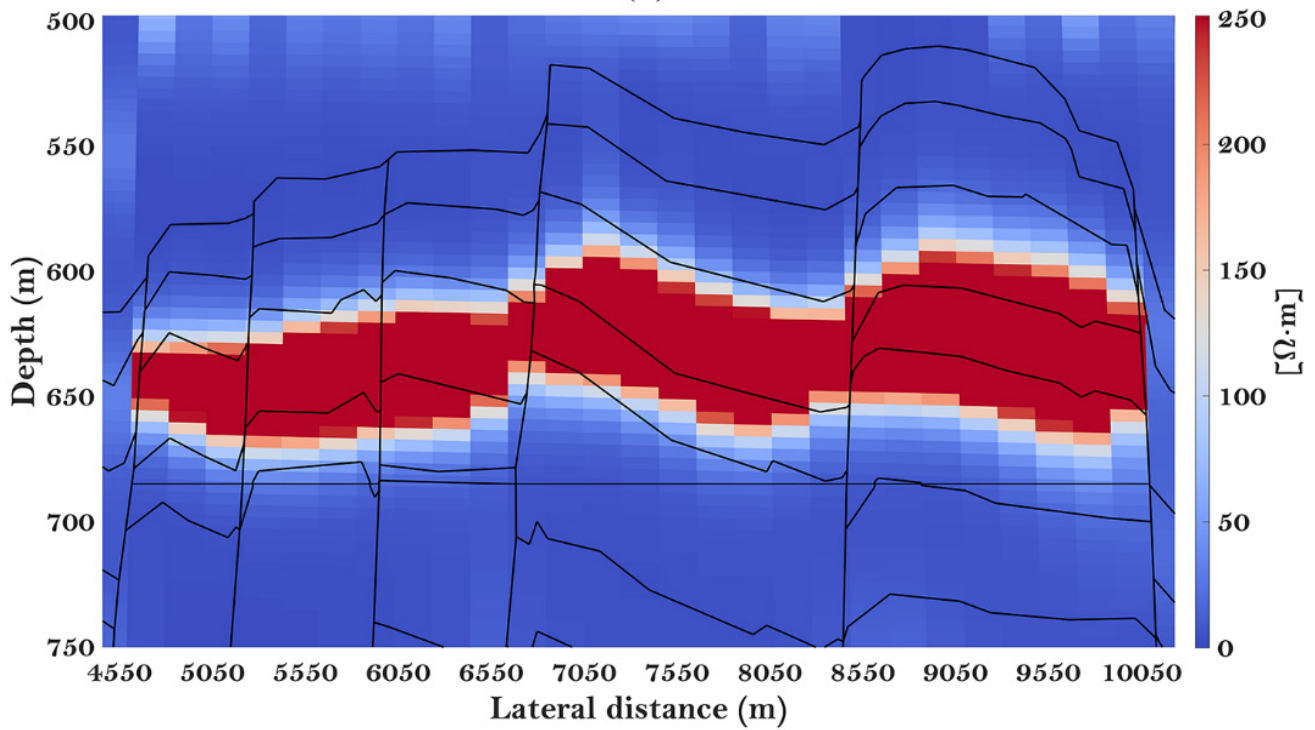

**Figure 3:** *(a) CSEM synthetic model of Wisting oil field (color coded with vertical resistivity). (b) Two-dimensional nonlinear inversion result of synthetic data generated by the wisting model.*

## 4.2 Synthetic PSFs

Examples of the PSF for six different locations within the interested domain are given in Figure 4. Notice that the PSF results shown in Figure 4 mean that PSF does not only vary in the field of view at certain location but also varies with the change of depth and lateral distance, and the shape of PSF is irregular, similar to traditional Gaussian function. It is also noted that some PSFs have negative side lobes, and we haven't determined the importance of negative side lobes, except that it shows that the way of PSFs average on adjacent parameters may be complicated, and it may not be symmetrical around the point of interest.

The PSFs in Figure 4c and 4e have obvious side lobes, and the corresponding model parameters of the PSFs are located at the edge of the reservoir. In addition, it can be seen from Table 3 that the offset distance between the maximum point of the PSFs and the center of the model parameter position are 40.00 m and 25.00 m, respectively. Based on the definition of PSF, these two PSFs have specific contributions to the geometrical distortion of the inverted model. Even the PSF in Figure 4a has no obvious side lobe, but the maximum value of the PSF lies off the center of corresponding model parameters, which will also cause geometrical distortion of the inverted model.

The center offset distance of the maximum value for PSFs in Figure 4b, 4d and 4f is 0, and its value ranks in the first three places among these six PSFs, indicating that the resolution of the nearby area is higher, but its value is not equal to the average value of the corresponding object sub-domain, so the final inversion result will have to face the problem of contrast deficiency (Friedel, 2003).

**Table 3:** *The position of maximum value for PSFs shown in Figure 4 and its Center offset distance from the location of model parameters.*

| Model parameter | Location of model parameter (m) | Maximum Values | Maximum point of the PSFs (m) | Center offset distance (m) |
|---|---|---|---|---|
| a | x = 6136.12, y = 613.91 | 0.0084 | x = 6136.12, y = 638.91 | 25.00 |
| b | x = 6134.27, y = 637.11 | 0.1173 | x = 6134.27, y = 637.11 | 0 |
| c | x = 7343.23, y = 578.20 | 0.0038 | x = 7343.23, y = 618.20 | 40.00 |
| d | x = 7346.00, y = 615.64 | 0.0982 | x = 7346.00, y = 615.64 | 0 |
| e | x = 8551.63, y = 598.49 | 0.0153 | x = 8551.63, y = 623.49 | 25.00 |
| f | x = 8558.02, y = 626.83 | 0.1697 | x = 8558.02, y = 626.83 | 0 |

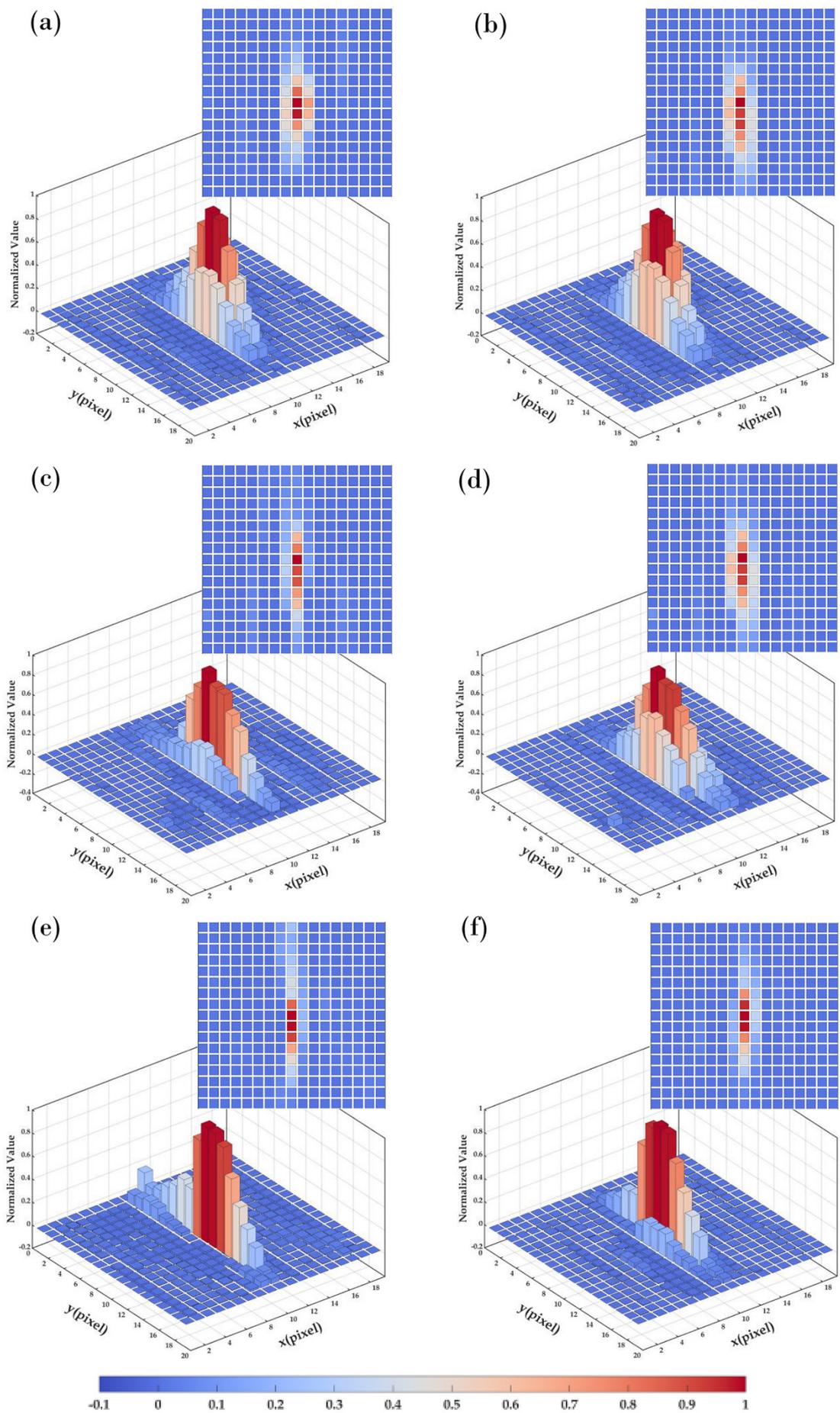
18

**Figure 4:** *The point spread functions at six different points within the imaging domain as determined from the model resolution matrix for the image given in Figure 3b. For presentation purposes the results have been normalized such that the maximum value in each case is equal to unity. (a) x = 6136.12 m, y = 613.91 m. (b) x = 6134.27 m, y = 637.11 m. (c) x = 7343.23 m, y = 578.20 m. (d) x = 7346.00 m, y = 615.64 m. (e) x = 8551.63 m, y = 598.49 m. (f) x= 8558.02 m, y = 626.83 m.*

## 4.3 Synthetic data deblurring

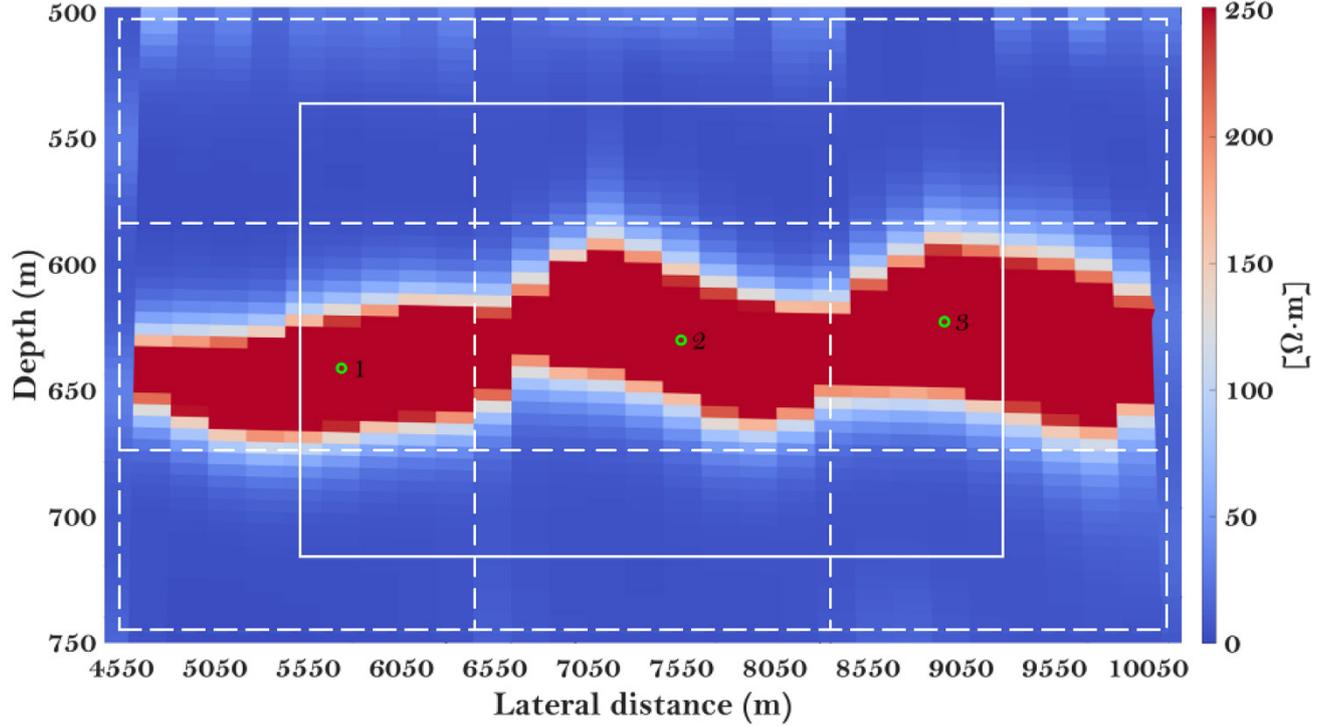

**Figure 5:** *The sectioning of the input image domain into 3 × 3 regions (area within the white dashed box). The domain section within the white solid frame for central sub-domain is overlaid to show the overlap of the PSF domain. Green circles indicates the position of sampled PSFs for the model parameters within the interested domain.*

In this section, the conducted experiment is reported and illustrated using a dataset of blurred image (Figure 3b) inverted from synthetic data. First of all, the interested area (the area within the white dashed box in Figure 5) was selected and then segmented into 3 × 3 sub-regions, as mentioned in Section 3.2, the partial overlap of the PSF domains has to be considered. In our experiment, we extend the disjoint sections to partially overlapping sections, the domain section within the white solid frame for the central sub-domain is overlaid to show the overlap of the PSF domains. Next, we sampled the PSFs (as shown in Figure 5, Green circles indicates the position of sampled PSFs for the model parameters) in each sub-domain within the interested domain, and a total of 3 PSFs were selected. For symmetrically displaying the PSF grid in the case of 3× 3, maximum point of each sampled PSF is taken as the center, and 6 pixels are selected around. After the outermost periphery is padded with zeros, a pixels size of 14 × 14 is obtained, as shown in Figure 6. It should be noted that here, six areas above and below the sub-areas numbered 1, 2 and 3 share the PSF information of the area between them. Further on those sampled PSFs were interpolated and the blurring matrix **A** was constructed, the blurred image can then be restored globally.

Regarding the parameters setting of NN-FCGLS, an initial guess of the iteration is set to zero vector, the noise level is not specified since how to estimate the noise level of blurred images has not been investigated in this paper, and the "safety factor" $\eta$ is set to be 1.01. The maximum number of inner iterations and the maximum number of outer iterations are set to 32 and 30 respectively, if the step length of CGLS update is less than 1e-15, it is considered that no update has occurred, and the inner iterations will be terminated. In addition, we set nonstop to force the iterations to continue beyond the number of iterations selected by the discrepancy principle stopping rule, hence the program will only stop running when the maximum number of total iterations reaches 100, and the results of each iteration will be saved. Here, Rnrm represents the relative residual norms at each iteration, $\|\mathbf{b} - \mathbf{A}x^{(k)}\|_2 / \|\mathbf{b}\|_2$, and NeRnrm represents the (penalized) normal equations residual norms at each iteration $\|\mathbf{A}^T(\mathbf{b} - \mathbf{A}x^{(k)})\|_2 / \|\mathbf{A}^T\mathbf{b}\|_2$, both of which are displayed in Figure 7. As can be seen from the Figure, the change of the residual norms has tended to be gentle after about 25 iterations, and the algorithm seems to have reached its convergence.

An analysis of the inversion results of iterative algorithms often involves a common problem, which is, over-fitting. It is impossible to solve all problems by only one convergence, that is, when the value of residual norms cannot be iterated to a smaller value, the solution obtained at this time could not be guaranteed to be the optimal solution, and it may be trapped in the local optimal solution. What is the best iteration number is not the topic of this paper, therefore, the best iteration is determined in this experiment by combining the convergence curves and the coincidence between the output results of several iterations and the synthetic model. There is no doubt that when we don't have any other verified models, geological data or logging data to prove the best result of deblurring, then, a tricky situation will need to be resolved.

The reconstruction results by using NN-FCGLS are presented in Figure 8a, where the increased resolution is obvious, and its iteration number is equal to 15, the corresponding values of Rnrm and NeRnrm shown in Figure 7 are 0.200 and 0.097, respectively. Due to pulse compression, we can easily see from Figure 8a that the vertical resolution has been enhanced, a large-scale smearing area outside the reservoir has been well eliminated. The latter has a higher resistivity value than the reservoir area of the synthetic model (Figure 3a), which demonstrated that the numerical accuracy of the restored model has also been well improved.

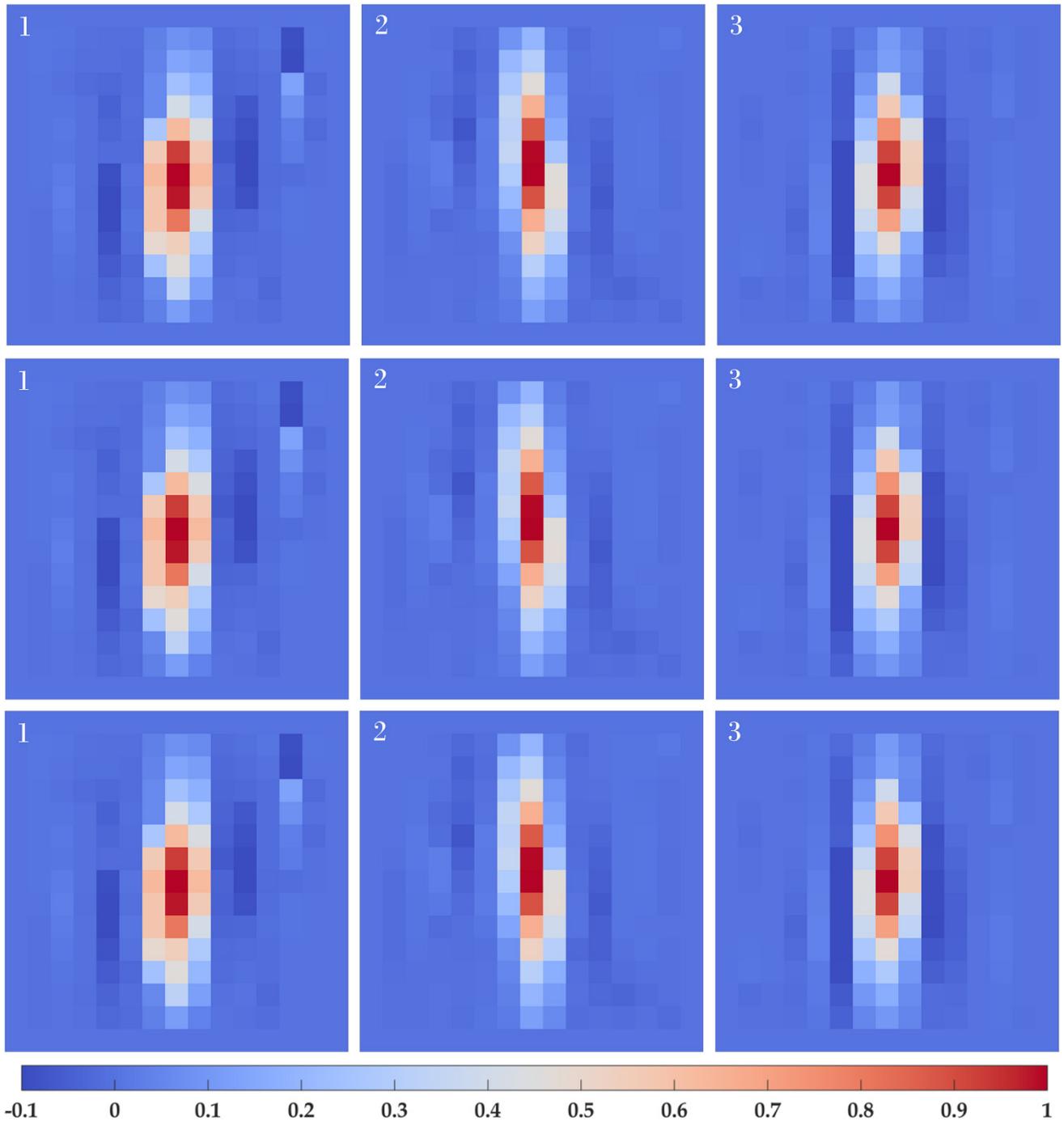

**Figure 6:** *The point spread functions at three different points within the imaging domain as determined from the model resolution matrix for the image given in Figure 3b. For presentation purposes the results have been normalized such that the maximum value in each case is equal to unity. (1) x= 5729.16 m, y = 641.09 m. (2) x= 7549.05 m, y = 630.00 m. (3) x = 8960.49 m, y = 622.69 m.*

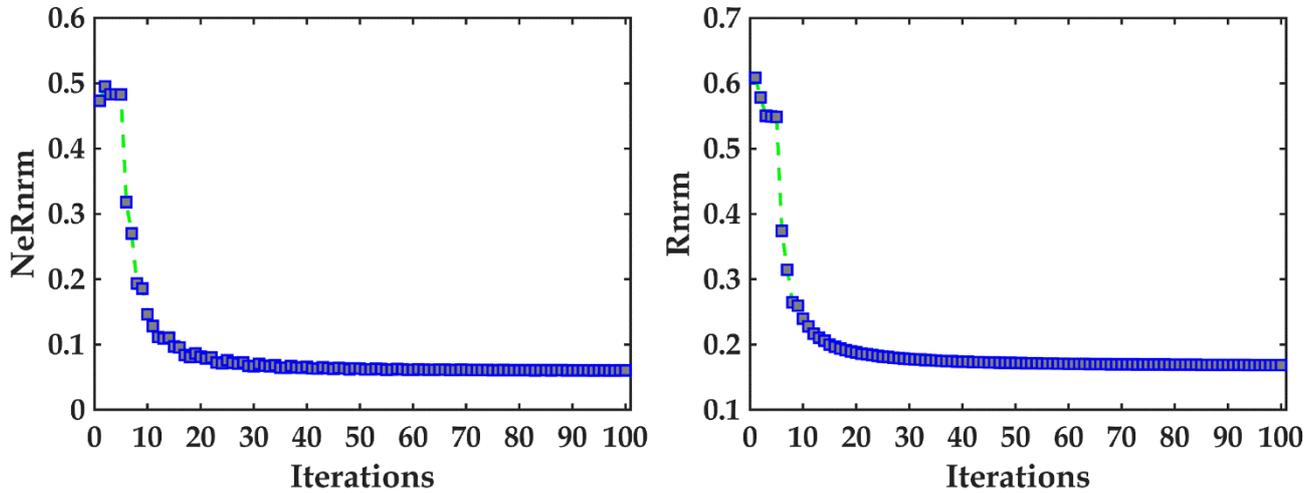

**Figure 7:** *Illustration of solution of the 2-D deblurring problem for the synthetic data. Left: Normal equations relative residual norms history and relative residual norms history at each iteration for the NN-FCGLS method.*

In addition, the reservoir position revealed by the blurred image (Figure 3b) seems to be more inclined to the above area of the abnormal area, because the resistivity of the bottom area (Depth between 625m and 660m) is closer to the actual value, and its contrast is more obvious. In fact, the reservoir is located in the stratum with a depth of 605m to 683m. The center of the reservoir revealed by blurred image and deblurred image is shallower than the actual position (left and middle compartment), but the thickness of the reservoir shown by deblurred image (Figure 8a) is closer to the actual value.

### 4.4  Result of Blind deconvolution for the synthetic data

The result of blind deconvolution using maximum likelihood estimation algorithm is shown in Figure 8b. Compared with the above result of spatially variable deblurring (Figure 8a), its vertical resolution is not as good as that of deconvolution using PSFs, and there are still many artifacts, but it is better than the blurred inversion result (Figure 3b). The challenge of blind deconvolution lies in the size of the array of initial estimated values of PSF and the determination of the weight matrix. In addition, the ringing effect caused by discrete Fourier transform used in the algorithm may appear in the output image.

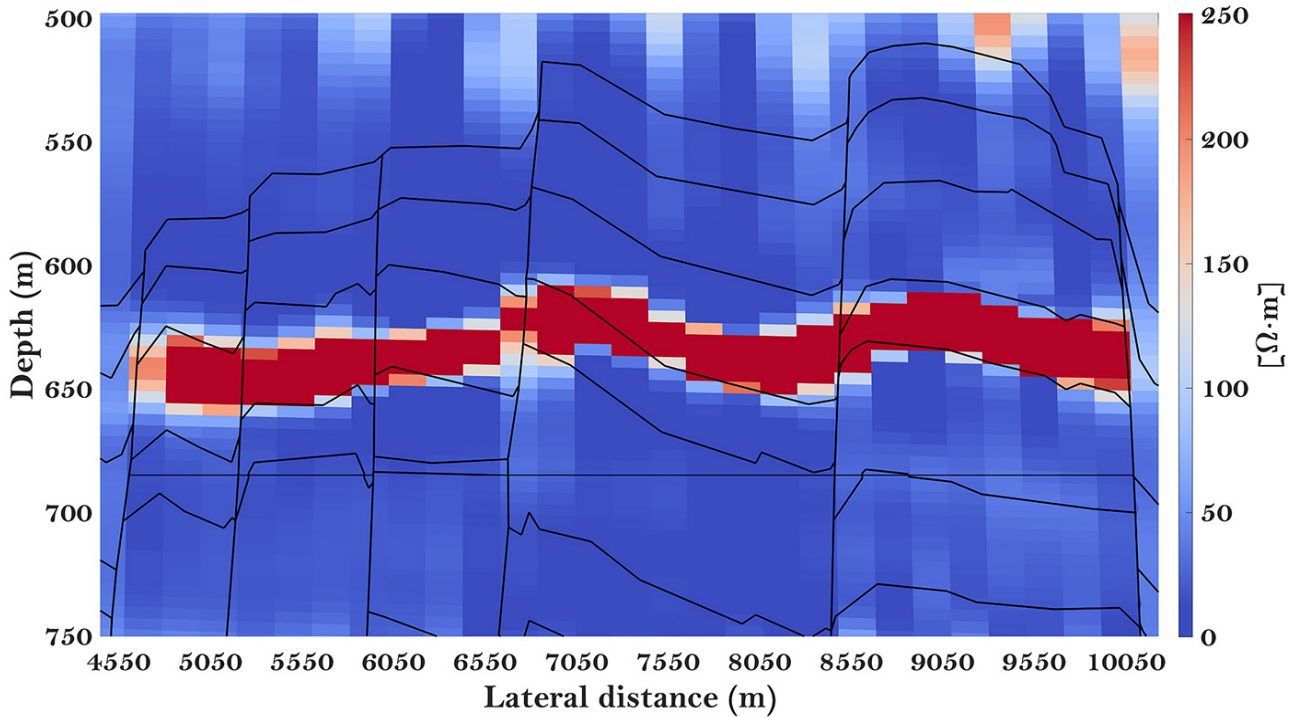

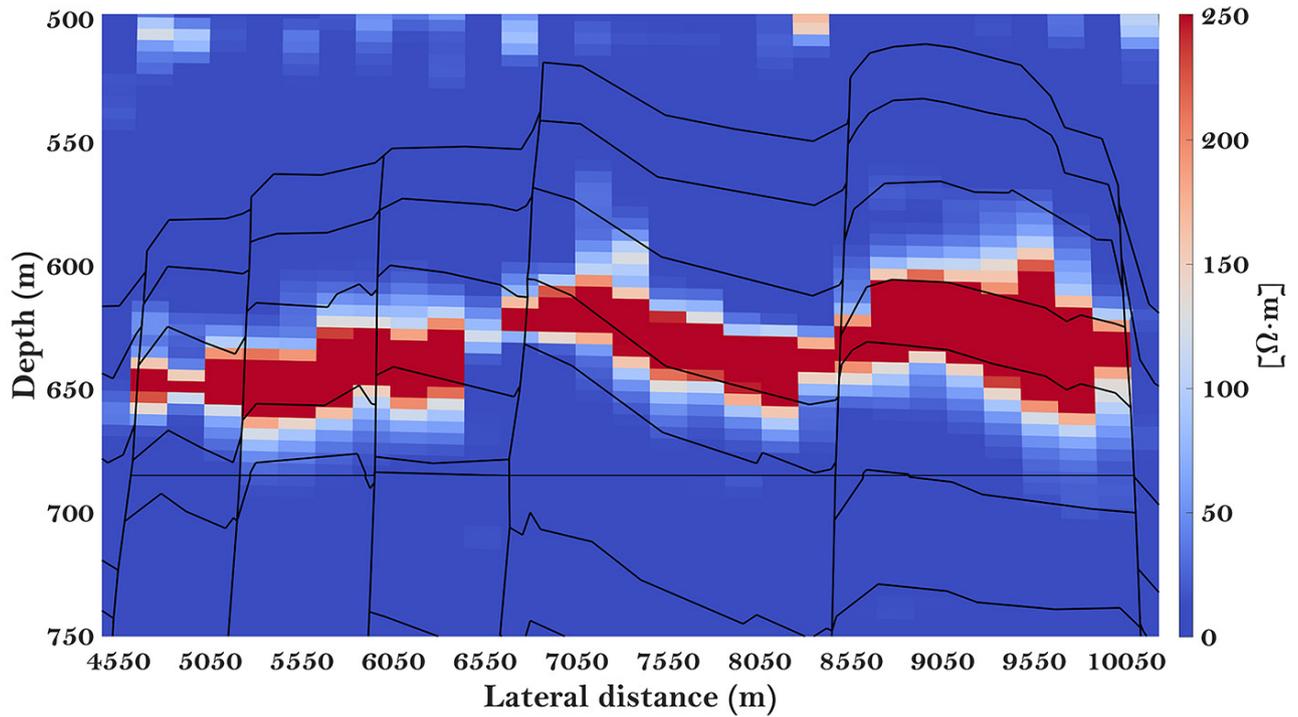

**Figure 8:** *(a) Space-Varing deblurred result of the Wisting CSEM synthetic data. (b) Blind deblurred result of the Wisting CSEM synthetic data.*

## 5 APPLICATION TO SURVEY DATA

### 5.1 Field data inversion

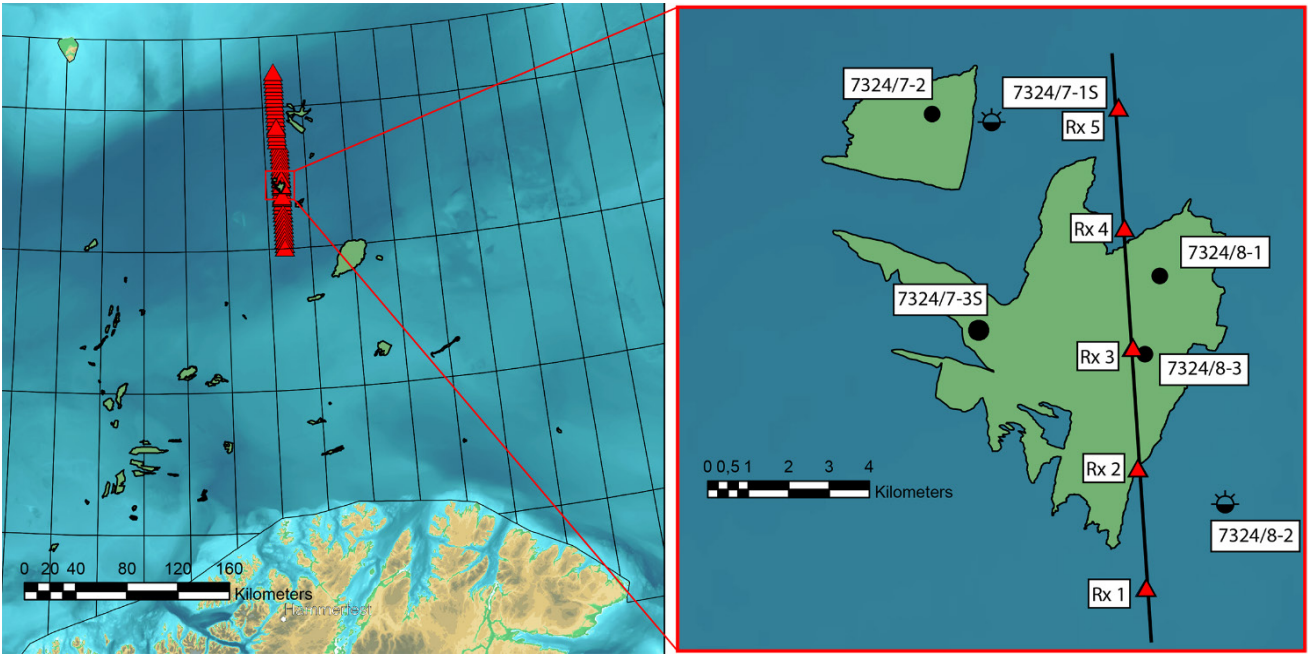

**Figure 9:** *Map of the South Western Barents Sea along with a zoomed section of the Wisting oil field. The selected receivers of the extracted 2D CSEM line are highlighted with red triangles, and its nearby wellbores named by numbers are also highlighted by black solid circles or black semicircles. This CSEM data was collected with a tow direction from south to north. Accordingly, data points where the source is south of its corresponding receiver are represented as in-tow ones, while the out-tow ones will have the source north of the corresponding receiver.*

To demonstrate the effectiveness of the proposed method on the field data, the preceding deconvolution technique is employed to CSEM field data of a 2D line, which is extracted from the BSMC08W 3D survey that belongs to the multiclient library of EMGS source. This data set were collected at the Wisting oil field, which is located in the Hoop fault complex north in the South Western Barents Sea. The right picture in Figure 9 highlights the corresponding selected receiver locations, and this survey was carried out during the summer of 2008. The synthetic model shown in Figure 3a is built from this field data combined with high-quality seismic data and resistivity logging data. The reservoir in Wisting is highly segmented and very shallow (250m below seabed), and contains oil in sandstone of Late Triassic age in the Fruholmen Formation and of Early Jurassic age in the Nordmela and Stø Formations.

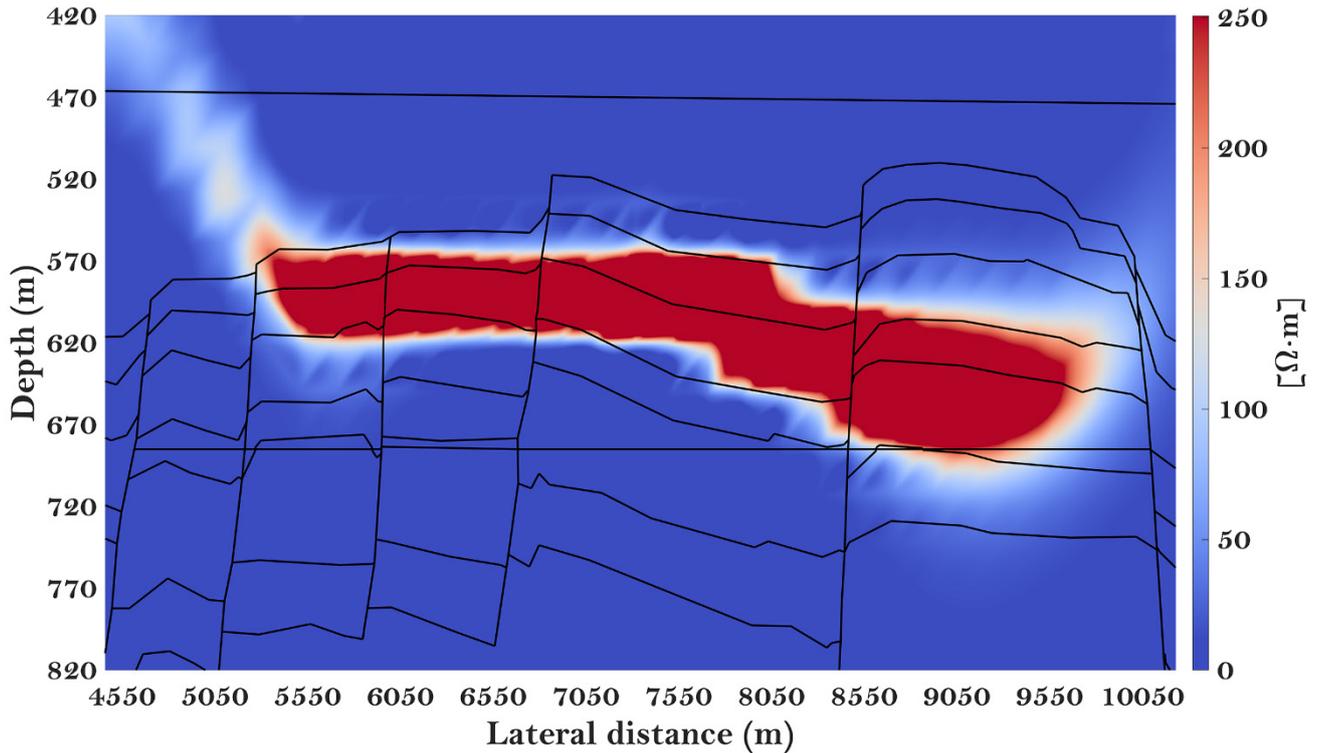

**Figure 10:** *Two-dimensional nonlinear inversion result of field data collected at the Wisting oil field in the South-Western Barents Sea.*

The full dataset up to 12 Hz was used as data input and the inversion results are shown in Figure 10. Direct comparison between the inverted result of the field data and that of the synthetic data suggests that the reservoir is appraised shallower in case of the former (Figures 3b and Figures 10). This phenomenon might be attributed to our scheme of employing a 2.5D inversion algorithm to deal with the 3D problem. We assume that there is no variations in the electrical properties along the strike direction when this 2.5-D technology is used. In addition, there is also an inherent hypothesis that the model extends to infinity along the identical direction. Therefore, the inversion algorithm attempts to compensate for these inconsistencies by positioning the reservoir shallower. Both the inversion results of field data and synthetic data show that the reservoir consists of three compartments. As shown, however, these three compartments are more distinctly split in the inversion results of synthetic data. In addition, the lateral extension of reservoir revealed by both cases is almost the same. A further notable feature of the inverted model of field data can be observed by looking at the rightmost compartment, which is positioned deeper in the model.

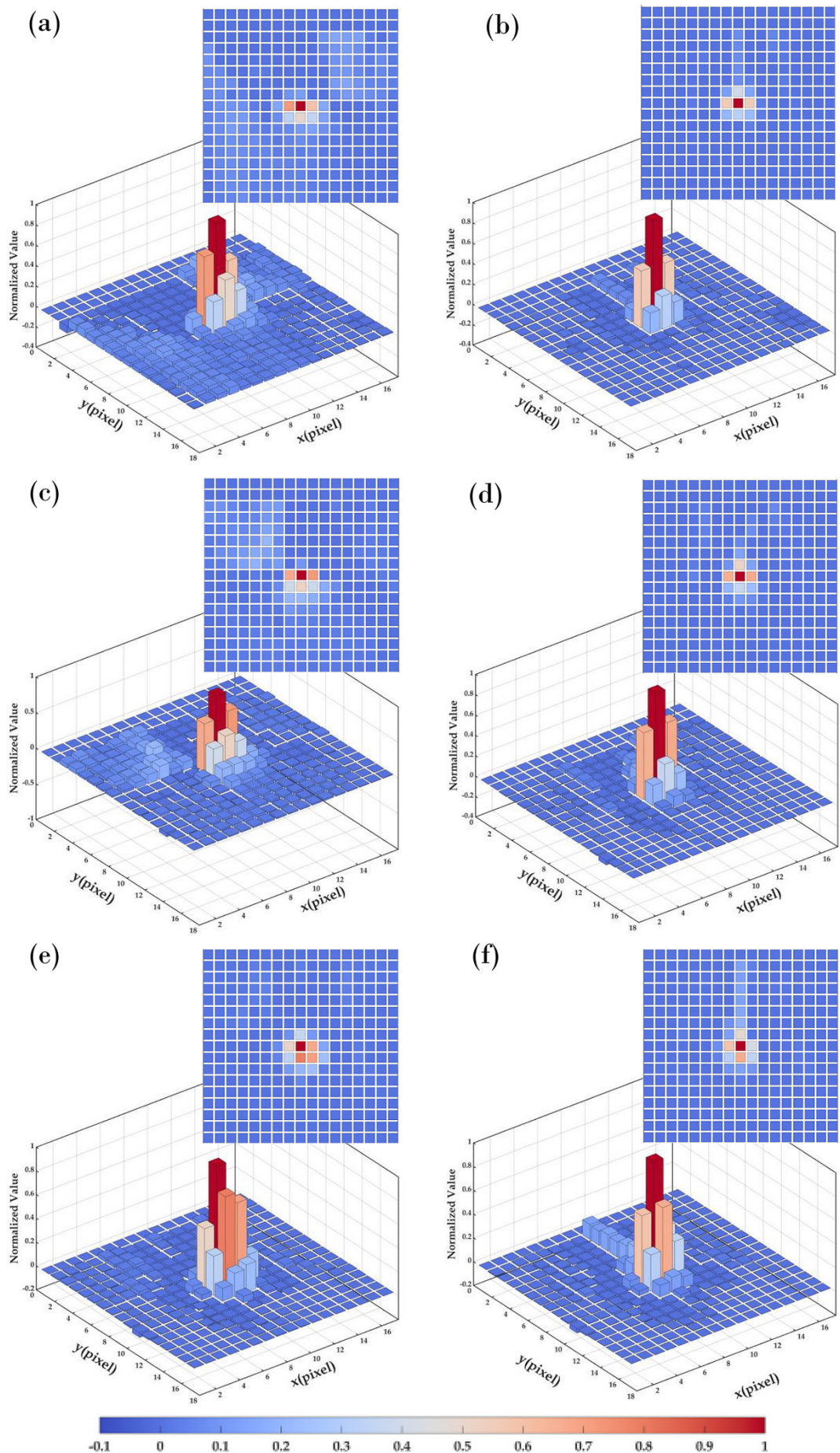

**Figure 11:** *The point spread functions at six different points within the imaging domain as determined from the model resolution matrix for the image given in Figure 10. For presentation purposes the results have been normalized such that the maximum value in each case is equal to unity. (a) x= 6100.01 m, y = 620.75 m. (b) x= 6100.02 m, y = 590.41 m. (c) x= 7300.02 m, y = 622.25 m. (d) x= 7300.02 m, y = 592.03 m. (e) x= 8500.02 m, y = 653.85 m. (f) x = 8500.02 m, y = 623.75 m.*

## 5.2   PSFs Calculated by Field data

The 6 PSFs calculated after the inversion of the field data are shown in Figure 11, and the specific positions of their corresponding model parameters can be found in Table 4. Compared with the synthetic PSFs, the PSFs calculated by field data has no obvious side lobe, even though three PSFs (Figure a, b and c) are located on the edge of the inverted model. The position of maximum value for PSFs shown in Figure 11 and its center offset distance from the location of model parameters are listed in Table 4, the characteristic similar to the synthetic PSFs is that, for PSFs whose maximum value is closer to 1 and located in the abnormal area with larger resistivity value, the distance between the maximum value point and the corresponding model parameter position is smaller.

**Table 4:** *The position of maximum value for PSFs shown in Figure 11 and its Center offset distance from the location of model parameters.*

| Model parameter | Location of model parameter (m) | Maximum Values | Maximum point of the PSFs (m) | Center offset distance (m) |
|---|---|---|---|---|
| a | x = 6100.01, y = 620.75 | 0.0354 | x = 6100.02, y = 590.41 | 30.34 |
| b | x = 6100.02, y = 590.41 | 0.4709 | x = 6100.02, y = 590.41 | 0 |
| c | x = 7300.02, y = 622.25 | 0.0192 | x = 7300.02, y = 592.02 | 29.85 |
| d | x = 7300.02, y = 592.03 | 0.3712 | x = 7300.02, y = 592.03 | 0 |
| e | x = 8500.02, y = 653.85 | 0.1702 | x = 8500.02, y = 623.75 | 30.10 |
| f | x = 8500.02, y = 623.75 | 0.3794 | x = 8500.02, y = 623.75 | 0 |

## 5.3   Field data deblurring

The results presented in the synthetic data deblurring demonstrate that it is effective and feasible to recover the final inverted model by extracting and employing the information contained in PSFs. The synthetic data, however, is associated with an ideal model case. Typically, in-situ data represent responses from more complex earth models, which are further complicated by the defects of measurement and instruments. We still believe that our proposed method is effective in real data. For supporting this statement, next, we give an example of the restoration of the inversion model for the field data acquired at Wisting oil field.

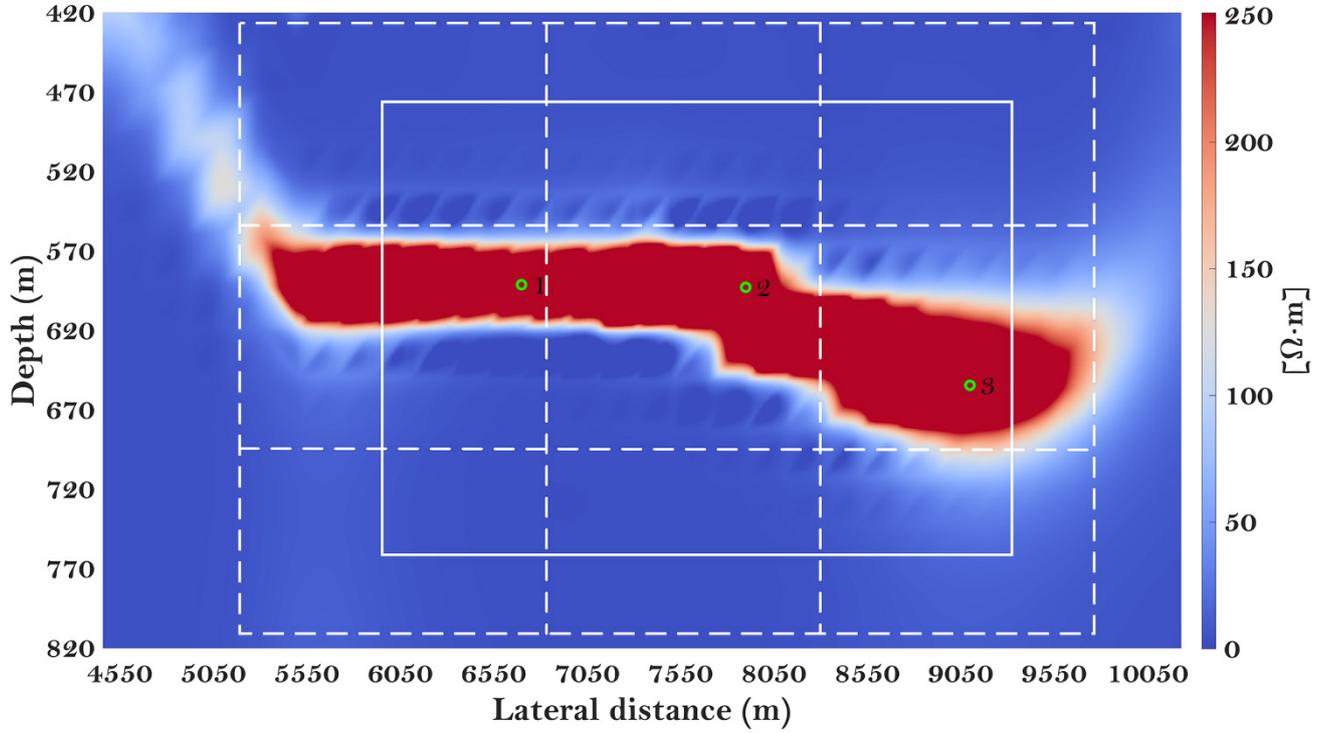

**Figure 12:** *The sectioning of the input image domain into 3 × 3 regions (area within the white dashed box). The domain section within the white solid frame for central sub-domain is overlaid to show the overlap of the PSF domains. Green circles indicates the position of sampled PSFs for the model parameters within the interested domain.*

As we did in the example of synthetic data deblurring, we first segmented a region of interest, as shown in Figure 12 of which the area within the white dashed box. Just like the example of synthetic data, the maximum values of our point spread function selected here are all located in the abnormal region shown by the inversion results (Figure 10), even though their corresponding model parameters may not be there. Three PSFs were selected, and the green circles show the positions of their corresponding model parameter. Due to the large mesh grid (200 m× 30 m) of the inversion model for the abnormal area in the case of the field data, we then interpolated the sampled PSFs, the generated PSFs are displayed in Figure 13 after a fixed number of pixels were selected around the maximum point for each PSF. Here, it should be noted that before using these interpolated PSFs to construct blurring matrix **A**, it is necessary to normalize the interpolated PSFs to ensure that the power of the interpolated

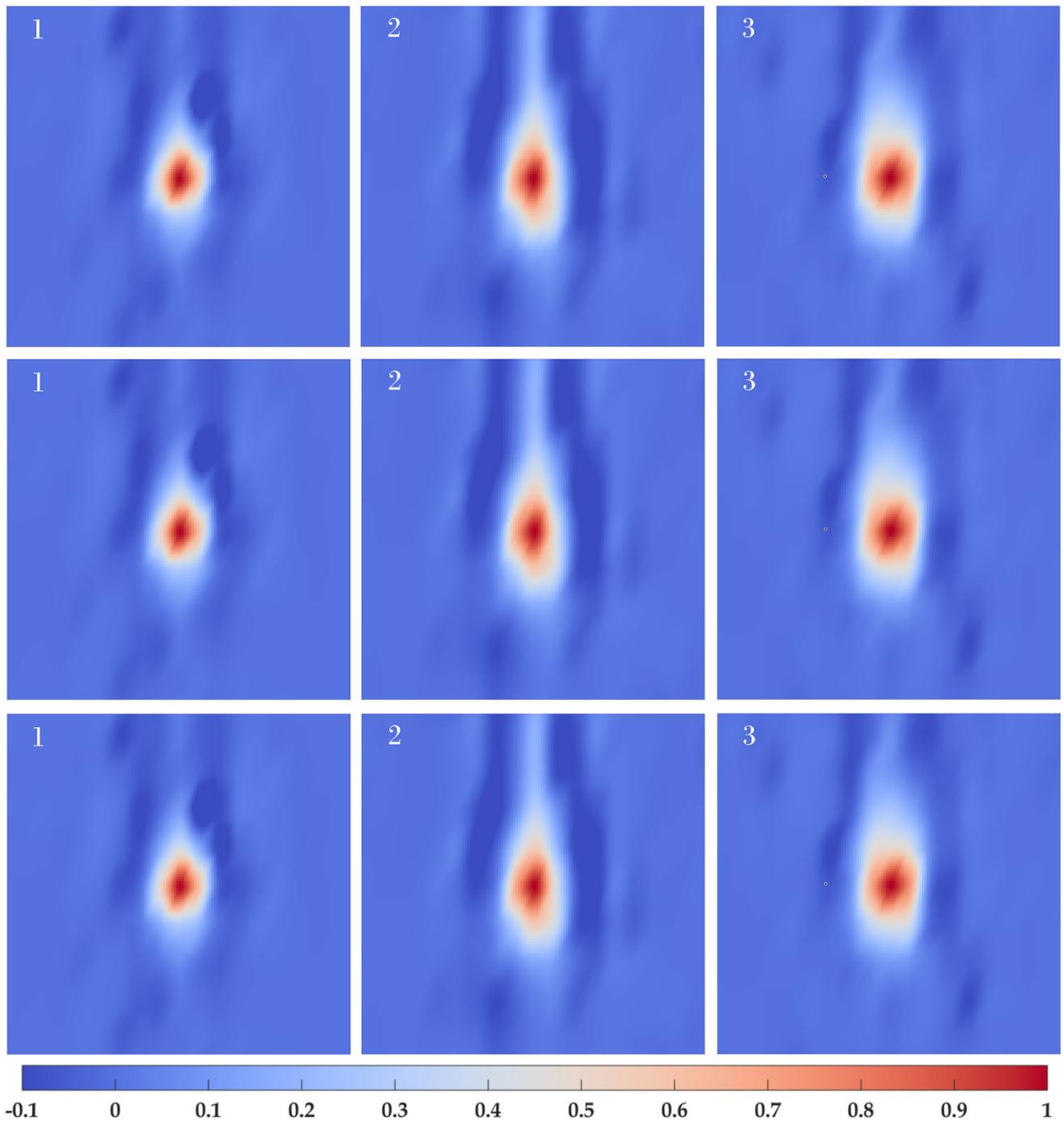

**Figure 13:** *The interpolated point spread functions at three different points within the imaging domain as determined from the model resolution matrix for the image given in Figure 10. For presentation purposes the results have been normalized such that the maximum value in each case is equal to unity. (1) x= 6700.02 m, y = 591.22 m. (2) x= 7900.02 m, y = 592.83 m. (3) x= 9100.02 m, y = 654.541iu1iujec m.*

PSFs remains unchanged, otherwise the pixel value of the output image will be far smaller than the actual value.

Regarding iterative parameter setting, an initial guess of the iteration is set to zero vector, the noise level is also not specified in this field case, and the "safety factor" $\eta$ is set to be 1.01. The maximum number of inner iterations and the maximum number of outer iterations are set to 32 and 30 respectively, and the nonstop was set to force the iterations only stop running when the maximum number of total iterations reaches 100.

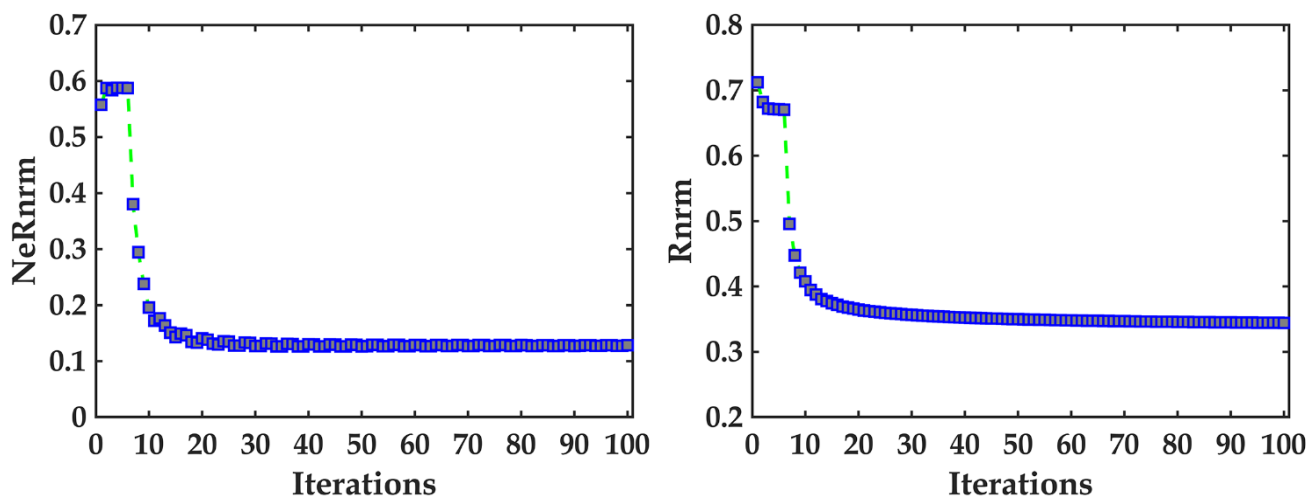

**Figure 14:** *Illustration of solution of the 2-D deblurring problem for the field data. Left: Normal equations relative residual norms history and relative residual norms history at each iteration for the NN-FCGLS method.*

The reconstruction results of the field data case by using NN-FCGLS are presented in Figure 15a, and its iteration number is equal to 16, the corresponding values of Rnrm and NeRnrm shown in Figure 14 are 0.364 and 0.133, respectively. The rightmost part of the original inversion model of the field data is rather vague. However, the resolution is greatly improved after deconvolution. The numerical accuracy of the recovery model is also improved, similar to the case of synthetic data. In addition, the reservoir thickness shown by the deblurring result also approaches the actual value.

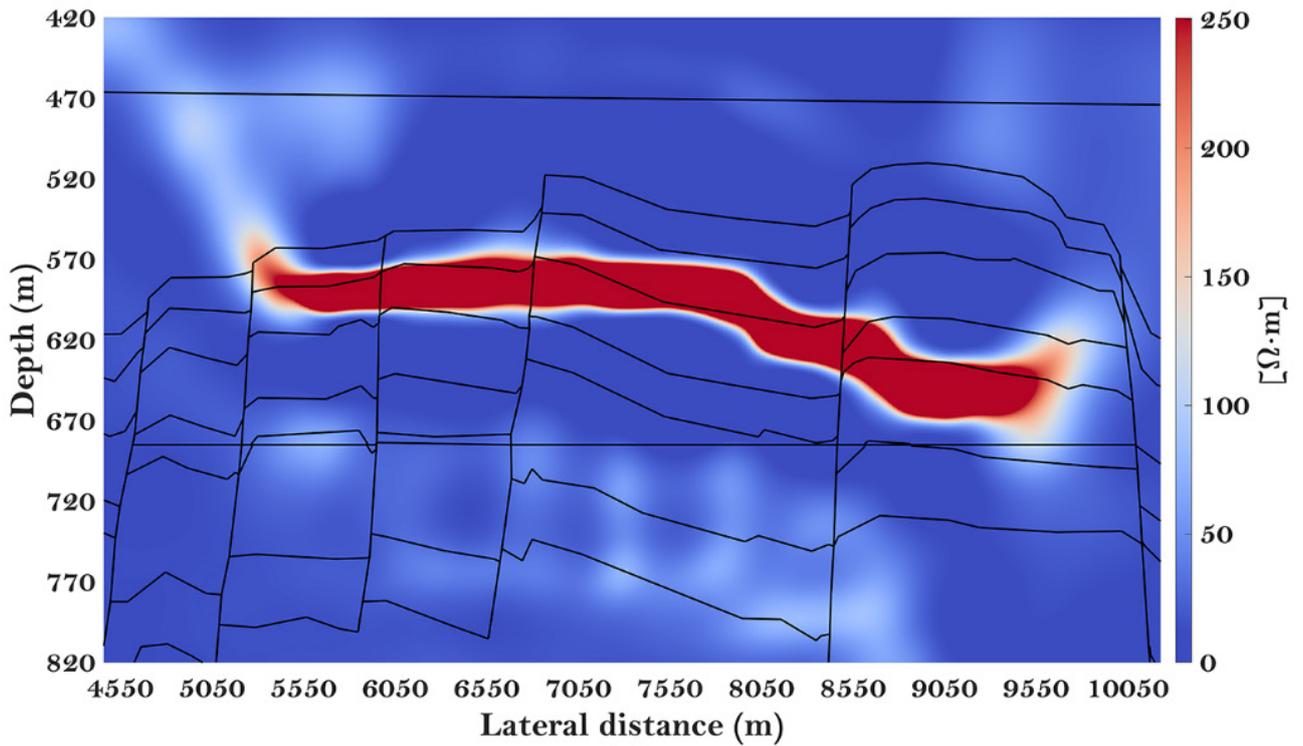

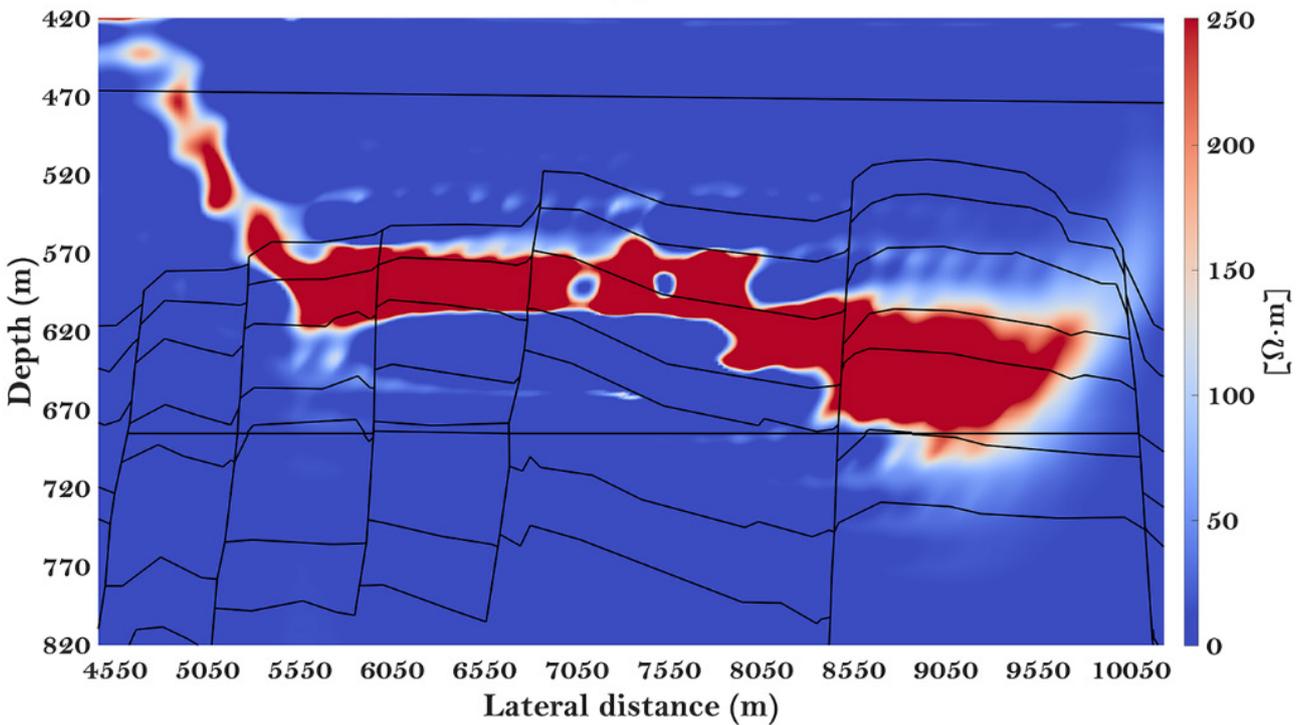

**Figure 15:** *(a) Deblurred result of the field data acquired at the Wisting oil field in the Barents Sea. (b) Blind deblurred result of the field data acquired at the Wisting oil field in the Barents Sea.*

### 5.4 Result of Blind deconvolution for the field data

The result of blind deconvolution using maximum likelihood estimation algorithm for field data is shown in Figure 15b. Compared with the above-mentioned result of space-varing deblurring (Figure 15a), the effect of blind deconvolution is not ideal. An obvious artifact appeared in the upper left corner, and the resolution of the rightmost part was blurred instead of improved. The blind deconvolution once again benchmarks our proposed approach.

## 6   CONCLUSION

In this paper, we have examined the possibility of the application of iterative restoration algorithms to improve the spatial resolution of output image by the inversion scheme. We calculate the resolution matrix associated with the inversion and derive the corresponding point spread functions (PSFs). The PSFs give information about how much the actual inversion has been blurred, and use of space-varying deconvolution can therefore further improve the inversion result. The PSFs vary in space, but existing computational methods for image restoration assume a spatially-invariant blurring. However, modeling the space-varying PSFs as a sequence of spatially invariant PSFs re-opens the possibility of reconstruction. Hence, this paper considers a space-varying blur model, which is based on the interpolation of the spatially invariant PSFs simulated for different sub-regions of the image domain. To study the efficiency of the blur model proposed, an iterative algorithms for solving a system of linear algebraic equations were implemented. The actual deblurring is carried out by use of the nonnegative flexible conjugate gradient algorithm for least squares problem (**NN-FCGLS**), which is a fast iterative restoration technique. In order to benchmark our proposed approach, we also introduce results obstained by use of a blind deconvolution algorithm based on maximum likelihood estimation (**MLE**) with unknown PSFs.

The potential of the proposed approach has been demonstrated using both complex synthetic data as well as field data acquired at the Wisting oil field in the Barents Sea. In both cases, the resolution of the final inversion result has improved and shows better agreement with the known target area, while not producing artifacts associated with large-amplitude events. The research on how different PSF combinations, iterative parameter settings, iterative convergence, deconvolution algorithms, etc., affect the resulting images is left for future papers.

**Author Contributions**

FP-L: Conceptualization, deconvolution methodology, restoration of 2-D inversion models, image Processing, Writing--First Draft, revision and editing. VS-T: Construction of Wisting synthetic model, inversion of CSEM synthetic and field data, calculation of resolution matrix and derive PSFs, revision. LJ-G: Led the writing of the manuscript, methodology, review and revision. JH-Y: Led the writing, conceptualization, review and revision of this manuscripts.

**Acknowledgments**

The authors acknowledge EMGS ASA for providing the field data and Dr. Kerry Key for making the MARE2DEM package publicly available. The authors also would like to thank Professor Silvia Gazzola, Per Christian Hansen, and James G. Nagy for willing to share their software. It is on the basis of their research work that the scheme described in this paper can then be implemented and improved more easily.

**Conflict of interest**

All authors in this article declare that the research was conducted in the absence of any commercial or financial relationships that could be construed as a potential conflict of interest.

# References

Ayers, G. R., & Dainty, J. C. (1988). Iterative blind deconvolution method and its applications. Optics letters, 13(7), 547-549.

Alumbaugh, D. L., & Morrison, H. F. (1995). Theoretical and practical considerations for crosswell electromagnetic tomography assuming a cylindrical geometry. Geophysics, 60(3), 846-870.

Alumbaugh, D. L., & Newman, G. A. (2000). Image appraisal for 2-D and 3-D electromagnetic inversion. Geophysics, 65(5), 1455-1467.

Anconelli, B., Bertero, M., Boccacci, P., Carbillet, M., & Lanteri, H. (2006). Reduction of boundary effects in multiple image deconvolution with an application to LBT LINC-NIRVANA. Astronomy & Astrophysics, 448(3), 1217-1224.

An, M. (2012). A simple method for determining the spatial resolution of a general inverse problem. Geophysical Journal International, 191(2), 849-864.

Al-Ameen, Z., & Sulong, G. (2015). Deblurring computed tomography medical images using a novel amended landweber algorithm. Interdisciplinary Sciences: Computational Life Sciences, 7(3), 319-325.

Backus, G., & Gilbert, F. (1968). The resolving power of gross earth data. Geophysical Journal International, 16(2), 169-205.

Backus, G., & Gilbert, F. (1970). Uniqueness in the inversion of inaccurate gross earth data. Philosophical Transactions of the Royal Society of London. Series A, Mathematical and Physical Sciences, 266(1173), 123-192.

Backus, G. (1970). Inference from inadequate and inaccurate data, I. Proceedings of the National Academy of Sciences, 65(1), 1-7.

Backus, G. (1970). Inference from inadequate and inaccurate data, II. Proceedings of the National Academy of Sciences, 65(2), 281-287.

Backus, G. (1970). Inference from inadequate and inaccurate data, III. Proceedings of the National Academy of Sciences, 67(1), 282-289.

Biggs, D. S., & Andrews, M. (1997). Acceleration of iterative image restoration algorithms. Applied optics, 36(8), 1766-1775.

Berryman, J. G. (2000). Analysis of approximate inverses in tomography I. Resolution analysis of common inverses. Optimization and Engineering, 1(1), 87-115.

Berryman, J. G. (2000). Analysis of approximate inverses in tomography II. Iterative inverses. Optimization and Engineering, 1(4), 437-473.

Bertero, M., & Boccacci, P. (2005). A simple method for the reduction of boundary effects in the Richardson-Lucy approach to image deconvolution. Astronomy & Astrophysics, 437(1), 369-374.

Brown, V., Hoversten, M., Key, K., & Chen, J. (2012). Resolution of reservoir scale electrical anisotropy from marine csem data. Geophysics, 77(2), E147–E158.

Bogiatzis, P., Ishii, M., & Davis, T. A. (2016). Towards using direct methods in seismic tomography: computation of the full resolution matrix using high-performance computing and sparse QR factorization. Geophysical Journal International, 205(2), 830-836.

Christensen-Dalsgaard, J., Hansen, P. C., & Thompson, M. J. (1993). Generalized singular value decomposition analysis of helioseismic inversions. Monthly Notices of the Royal Astronomical Society, 264(3), 541-564.

Constable, S. (2010). Ten years of marine CSEM for hydrocarbon exploration. *Geophysics*, 75(5), 75A67-75A81.

Chrapkiewicz, K., Wilde-Piórko, M., Polkowski, M., & Grad, M. (2020). Reliable workflow for inversion of seismic receiver function and surface wave dispersion data: a "13 BB Star" case study. Journal of Seismology, 24(1), 101-120.

Dosso, S. E., & Oldenburg, D. W. (1989). Linear and non-linear appraisal using extremal models of bounded variation. Geophysical Journal International, 99(3), 483-495.

Dosso, S. E., & Oldenburg, D. W. (1991). Magnetotelluric appraisal using simulated annealing. Geophysical Journal International, 106(2), 379-385.

Du, Z., Querendez, E., & Jordan, M. (2012, June). Resolution and uncertainty in 3D stereotomographic inversion. In 74th EAGE Conference and Exhibition incorporating EUROPEC 2012 (pp. cp-293). European Association of Geoscientists & Engineers.

Ellingsrud, S., Eidesmo, T., Johansen, S., Sinha, M. C., MacGregor, L. M., & Constable, S. (2002). Remote sensing of hydrocarbon layers by seabed logging (SBL): Results from a cruise offshore Angola. *The Leading Edge*, 21(10), 972-982.

Friedel, S. (2003). Resolution, stability and efficiency of resistivity tomography estimated from a generalized inverse approach. Geophysical Journal International, 153(2), 305-316.

Fichtner, A., & Leeuwen, T. V. (2015). Resolution analysis by random probing. Journal of Geophysical Research: Solid Earth, 120(8), 5549-5573.

Gouveia, W. P., & Scales, J. A. (1997). Resolution of seismic waveform inversion: Bayes versus Occam. Inverse problems, 13(2), 323.

Günther, T. (2004). Inversion methods and resolution analysis for the 2D/3D reconstruction of resistivity structures from DC measurements. PhD thesis, TU Bergakademie Freiberg.

Gao, G., Alumbaugh, D., Chen, J., & Eyl, K. (2007, September). Resolution and uncertainty analysis for marine CSEM and cross-well EM imaging. In 2007 SEG Annual Meeting. OnePetro.

Golub, G. H., & Van Loan, C. F. (2013). Matrix computations. Johns Hopkins University Press, Baltimore.

Grayver, A. V., Streich, R., & Ritter, O. (2014). 3D inversion and resolution analysis of land-based CSEM data from the Ketzin $CO_2$ storage formation. Geophysics, 79(2), E101-E114.

Gazzola, S., & Wiaux, Y. (2017). Fast nonnegative least squares through flexible Krylov subspaces. SIAM Journal on Scientific Computing, 39(2), A655-A679.

Gazzola, S., Hansen, P. C., & Nagy, J. G. (2019). IR Tools: a MATLAB package of iterative regularization methods and large-scale test problems. Numerical Algorithms, 81(3), 773-811.

Hansen, P. C. (1990). Truncated singular value decomposition solutions to discrete ill-posed problems with ill-determined numerical rank. SIAM Journal on Scientific and Statistical Computing, 11(3), 503-518.

Holmes, T. J. (1992). Blind deconvolution of quantum-limited incoherent imagery: maximum-likelihood approach. JOSA A, 9(7), 1052-1061.

Holmes, T. J., Bhattacharyya, S., Cooper, J. A., Hanzel, D., Krishnamurthi, V., Lin, W. C., & Turner, J. N. (1995). Light microscopic images reconstructed by maximum likelihood deconvolution. In Handbook of biological confocal microscopy (pp. 389-402). Springer, Boston, MA.

Holmes, T. J., Biggs, D., & Abu-Tarif, A. (2006). Blind deconvolution. In Handbook of biological confocal microscopy (pp. 468-487). Springer, Boston, MA.

Hoversten, G. M., Røsten, T., Hokstad, K., Alumbaugh, D., Horne, S., & Newman, G. A. (2006). Integration of multiple electromagnetic imaging and inversion techniques for prospect evaluation. In SEG Technical Program Expanded Abstracts 2006 (pp. 719–723). Society of Exploration Geophysicists.

Jackson, D. D. (1972). Interpretation of inaccurate, insufficient and inconsistent data. Geophysical Journal International, 28(2), 97-109.

Jackson, D. D. (1979). The use of a priori data to resolve non-uniqueness in linear inversion. Geophysical Journal International, 57(1), 137-157.

Krishnamurthi, V., Liu, Y. H., Bhattacharyya, S., Turner, J. N., & Holmes, T. J. (1995). Blind deconvolution of fluorescence micrographs by maximum-likelihood estimation. Applied optics, 34(29), 6633-6647.

Keating, P. B. (1998). Weighted Euler deconvolution of gravity data. Geophysics, 63(5), 1595-1603.

Kalscheuer, T., De los Ángeles García Juanatey, M., Meqbel, N., & Pedersen, L. B. (2010). Non-linear model error and resolution properties from two-dimensional single and joint inversions of direct current resistivity and radiomagnetotelluric data. Geophysical Journal International, 182(3), 1174-1188.

Kazemi, N. (2018). Automatic blind deconvolution with Toeplitz-structured sparse total least squares Toeplitz-structured sparse TLS. Geophysics, 83(6), V345-V357.

Li, X. P., Söllner, W., & Hubral, P. (1995). Elimination of harmonic distortion in vibroseis data. Geophysics, 60(2), 503-516.

Li, X. P. (1997). Elimination of ghost noise in vibroseis data by deconvolution. Geophysical prospecting, 45(6), 909-929.

Lu, X. & Xia, C. (2007). Understanding anisotropy in marine CSEM data. In 2007 SEG Annual Meeting: OnePetro.

La Camera, A., Schreiber, L., Diolaiti, E., Boccacci, P., Bertero, M., Bellazzini, M., & Ciliegi, P. (2015). A method for space-variant deblurring with application to adaptive optics imaging in astronomy. Astronomy & Astrophysics, 579, A1.

Miller, C. R., & Routh, P. S. (2007). Resolution analysis of geophysical images: Comparison between point spread function and region of data influence measures. Geophysical Prospecting, 55(6), 835-852.

Menke, W. (1984). The resolving power of cross-borehole tomography. Geophysical Research Letters, 11(2), 105-108.

Menke, W. (1989) Geophysical Data Analysis: Discrete Inverse Theory. 1st Edition, Academic Press, San Diego, 289.

Menke, W. (2012). Geophysical data analysis: discrete inverse theory: MATLAB edition (Vol. 45). Academic press.

Mordret, A., Landès, M., Shapiro, N. M., Singh, S. C., Roux, P., & Barkved, O. I. (2013). Near-surface study at the Valhall oil field from ambient noise surface wave tomography. Geophysical Journal International, 193(3), 1627-1643.

Mattsson, J. (2015, October). Resolution and Precision of Resistivity Models from inversion of Towed Streamer EM data. In 2015 SEG Annual Meeting. OnePetro.

Nagy, J. G., & O'leary, D. P. (1997, October). Fast iterative image restoration with a spatially varying PSF. In Advanced Signal Processing: Algorithms, Architectures, and Implementations VII (Vol. 3162, pp. 388-399). SPIE.

Nagy, J. G., & O'Leary, D. P. (1998). Restoring images degraded by spatially variant blur. SIAM Journal on Scientific Computing, 19(4), 1063-1082.

Nagy, J. G., Palmer, K., & Perrone, L. (2004). Iterative methods for image deblurring: a Matlab object-oriented approach. Numerical Algorithms, 36(1), 73-93.

Newman, G. A., Commer, M., & Carazzone, J. J. (2010). Imaging CSEM data in the presence of electrical anisotropy. Geophysics, 75(2), F51–F61.

Oldenburg, D. W. (1983). Funnel functions in linear and nonlinear appraisal. Journal of Geophysical Research: Solid Earth, 88(B9), 7387-7398.

Oldenborger, G. A., & Routh, P. S. (2009). The point-spread function measure of resolution for the 3-D electrical resistivity experiment. Geophysical Journal International, 176(2), 405-414.

Robinson, E. A. (1967). Predictive decomposition of time series with application to seismic exploration. Geophysics, 32(3), 418-484.

Reid, A. B., Allsop, J. M., Granser, H., Millett, A. T., & Somerton, I. W. (1990). Magnetic interpretation in three dimensions using Euler deconvolution. Geophysics, 55(1), 80-91.

Routh, P. S., Oldenborger, G. A., & Wang, D. W. (2005, November). Optimal survey design using the point spread function measure of resolution. In 2005 SEG Annual Meeting. OnePetro.

Routh, P. S., & Miller, C. R. (2006, January). Image interpretation using appraisal analysis. In Symposium on the Application of Geophysics to Engineering and Environmental Problems 2006 (pp. 1812-1820). Society of Exploration Geophysicists.

Routh, P. S. (2009). A practical strategy to interrogate resolution, uncertainty and value of information in geophysical inverse problem. GEOHORIZONS, 14(2), 27-37.

Ren, Z., & Kalscheuer, T. (2020). Uncertainty and resolution analysis of 2D and 3D inversion models computed from geophysical electromagnetic data. Surveys in Geophysics, 41(1), 47-112.

Sen, M. K., Bhattacharya, B. B., & Stoffa, P. L. (1993). Nonlinear inversion of resistivity sounding data. Geophysics, 58(4), 496-507.

Spies, B. R., & Habashy, T. M. (1995). Sensitivity analysis of crosswell electromagnetics. Geophysics, 60(3), 834-845.

Sjoeberg, T. A., Gelius, L. J., & Lecomte, I. (2003, October). 2-D deconvolution of seismic image blur. In 2003 SEG Annual Meeting. OnePetro.

Senger, K., Birchall, T., Betlem, P., Ogata, K., Ohm, S., Olaussen, S., & Paulsen, R. S. (2021). Resistivity of reservoir sandstones and organic rich shales on the barents shelf: Implications for interpreting csem data. Geoscience Frontiers, 12(6),101063.

Torres–Verdin, C., (1991), Continuous profiling of magnetotelluric fields: Ph.D. dissertation, Univ. California, Berkeley.

Um, E. S., Alumbaugh, D. L. (2007). On the physics of the marine controlled-source electromagnetic method. *Geophysics*, 72(2), WA13-WA26.

Vio, R., Nagy, J., Tenorio, L., & Wamsteker, W. (2004). A simple but efficient algorithm for multiple-image deblurring. Astronomy & Astrophysics, 416(1), 403-410.

Wang, L., Zhao, Q., Gao, J., Xu, Z., Fehler, M., & Jiang, X. (2016). Seismic sparse-spike deconvolution via Toeplitz-sparse matrix factorization. Geophysics, 81(3), V169-V182.

Xia, J., Miller, R. D., & Xu, Y. (2008). Data-resolution matrix and model-resolution matrix for Rayleigh-wave inversion using a damped least-squares method. Pure and Applied Geophysics, 165(7), 1227-1248.

Yang, J., Huang, J., Zhu, H., McMechan, G., & Li, Z. (2022). An efficient and stable high-resolution seismic imaging method: Point-spread function deconvolution. Journal of Geophysical Research: Solid Earth, 127, e2021JB023281.

Zhou, Q., Becker, A., & Morrison, H. F. (1993). Audio-frequency electromagnetic tomography in 2-D. Geophysics, 58(4), 482-495.

Zuo, B., & Hu, X. (2012). Geophysical model enhancement technique based on blind deconvolution. Computers & Geosciences, 49, 170-181.